\documentclass[twocolumn]{aastex631}

\usepackage{xspace} 
\usepackage{upgreek} 
\usepackage{subfig}
\newcommand{\um}{$\upmu$m\xspace}	

\begin{document}

\title{The WINTER Observatory: \\ A One-Degree InGaAs Survey Camera to study the Transient Infrared Sky}

\author[0000-0002-7197-9004]{Danielle Frostig}
\affil{Center for Astrophysics $\vert$ Harvard \& Smithsonian, 60 Garden Street, Cambridge, MA 02138, USA}
\affiliation{Department of Physics and Kavli Institute for Astrophysics and Space Research, Massachusetts Institute of Technology, 77 Massachusetts
Ave, Cambridge, MA 02139, USA} 
\email{danielle.frostig@cfa.harvard.edu}

\author[0000-0002-4585-9981]{Nathan~P.~Lourie}
\affiliation{Department of Physics and Kavli Institute for Astrophysics and Space Research, Massachusetts Institute of Technology, 77 Massachusetts
Ave, Cambridge, MA 02139, USA} 

\author[0000-0003-2758-159X]{Viraj~Karambelkar}
\affil{Cahill Center for Astrophysics, California Institute of Technology, Pasadena, CA 91125, USA}
\affil{Department of Astronomy and Columbia Astrophysics Laboratory, Columbia University, 550 W 120th St. MC 5246, New York, NY 10027, USA}

\author[0000-0002-5619-4938]{Mansi~M.~Kasliwal}
\affil{Cahill Center for Astrophysics, California Institute of Technology, Pasadena, CA 91125, USA}

\author{Andrew~Malonis}
\affiliation{Department of Physics and Kavli Institute for Astrophysics and Space Research, Massachusetts Institute of Technology, 77 Massachusetts
Ave, Cambridge, MA 02139, USA} 

\author[0000-0003-3769-9559]{Robert A. Simcoe}
\affiliation{Department of Physics and Kavli Institute for Astrophysics and Space Research, Massachusetts Institute of Technology, 77 Massachusetts
Ave, Cambridge, MA 02139, USA} 

\author[0000-0003-2434-0387]{Robert~Stein}
\affil{Cahill Center for Astrophysics, California Institute of Technology, Pasadena, CA 91125, USA}
\affil{Department of Astronomy, University of Maryland, College Park, MD 20742, USA}
\affil{Joint Space-Science Institute, University of Maryland, College Park, MD 20742, USA} 
\affil{Astrophysics Science Division, NASA Goddard Space Flight Center, Mail Code 661, Greenbelt, MD 20771, USA}

\author{John W. Baker}
\affil{Caltech Optical Observatories, California Institute of Technology, 1216 E California Blvd, Pasadena, CA 91125, USA}

\author{Kevin~Burdge}
\affiliation{Department of Physics and Kavli Institute for Astrophysics and Space Research, Massachusetts Institute of Technology, 77 Massachusetts
Ave, Cambridge, MA 02139, USA} 

\author{Rick Burruss}
\affil{Caltech Optical Observatories, California Institute of Technology, 1216 E California Blvd, Pasadena, CA 91125, USA}

\author{Curt Corcoran}
\affil{Caltech Optical Observatories, California Institute of Technology, 1216 E California Blvd, Pasadena, CA 91125, USA}

\author[0000-0002-8989-0542]{Kishalay~De}
\affiliation{Department of Physics and Kavli Institute for Astrophysics and Space Research, Massachusetts Institute of Technology, 77 Massachusetts
Ave, Cambridge, MA 02139, USA} 
\affil{Department of Astronomy and Columbia Astrophysics Laboratory, Columbia University, 550 W 120th St. MC 5246, New York, NY 10027, USA}
\affil{Center for Computational Astrophysics, Flatiron Institute, 162 5th Ave., New York, NY 10010, USA}

\author[0000-0001-8467-9767]{G{\'a}bor~F{\H u}r{\'e}sz}
\affiliation{Department of Physics and Kavli Institute for Astrophysics and Space Research, Massachusetts Institute of Technology, 77 Massachusetts Ave, Cambridge, MA 02139, USA}

\author{Nicolae Ganciu}
\affil{Caltech Optical Observatories, California Institute of Technology, 1216 E California Blvd, Pasadena, CA 91125, USA}

\author{Kari Haworth}
\affil{Center for Astrophysics $\vert$ Harvard \& Smithsonian, 60 Garden Street, Cambridge, MA 02138, USA}

\author{Carolyn M. Heffner}
\affil{Caltech Optical Observatories, California Institute of Technology, 1216 E California Blvd, Pasadena, CA 91125, USA}

\author{Erik Hinrichsen}
\affiliation{Department of Physics and Kavli Institute for Astrophysics and Space Research, Massachusetts Institute of Technology, 77 Massachusetts
Ave, Cambridge, MA 02139, USA} 

\author[0000-0001-5926-3911]{Jill~Juneau}
\affiliation{Department of Physics and Kavli Institute for Astrophysics and Space Research, Massachusetts Institute of Technology, 77 Massachusetts
Ave, Cambridge, MA 02139, USA}

\author[0000-0001-6331-112X]{Geoffrey~Mo}
\affiliation{Department of Physics and Kavli Institute for Astrophysics and Space Research, Massachusetts Institute of Technology, 77 Massachusetts
Ave, Cambridge, MA 02139, USA} 
\affil{Department of Astronomy, California Institute of Technology, 1216 E California Blvd, Pasadena, CA 91125, USA}
\affil{Observatories of the Carnegie Institution of Washington, 813 Santa Barbara St, Pasadena, CA 91101, USA}

\author[0000-0003-1227-3738]{Josiah Purdum}
\affil{Cahill Center for Astrophysics, California Institute of Technology, Pasadena, CA 91125, USA}

\author[0000-0003-4725-4481]{Sam Rose}
\affil{Cahill Center for Astrophysics, California Institute of Technology, Pasadena, CA 91125, USA}

\author{Cruz~Soto}
\affiliation{Department of Physics and Kavli Institute for Astrophysics and Space Research, Massachusetts Institute of Technology, 77 Massachusetts
Ave, Cambridge, MA 02139, USA} 
\affiliation{Stanford University School of Engineering, 475 Via Ortega, Fl 3, Stanford, CA 94305, USA} 

\author{Jeffry Zolkower}
\affil{Caltech Optical Observatories, California Institute of Technology, 1216 E California Blvd, Pasadena, CA 91125, USA}

\begin{abstract}

The Wide-field Infrared Transient Explorer (WINTER) is a near-infrared time-domain survey instrument operating on a dedicated 1-meter robotic telescope at Palomar Observatory. The project takes advantage of recent technology advances in time-domain astronomy, robotic telescopes, large-format sensors, and rapid data reduction and alert software for timely follow up of events. Since June of 2023, WINTER robotically surveys the sky each night to a median depth of $J_{\text AB} = 18.5$ mag, balancing a variety of science programs including searching for kilonovae from gravitational-wave alerts, blind surveys to study galactic and extragalactic transients and variables, and building up reference images of the near-infrared sky. The project also serves as a technology demonstration for new large-format Indium Gallium Arsenide (InGaAs) sensors for wide-field science in the near infrared without cryogenically cooled optics or detectors. WINTER’s custom camera combines six InGaAs sensors with a novel tiled fly’s-eye optical design to cover a $\textgreater$1 deg$^2$ field of view with 1 arcsecond pixels in the $Y$-, $J$-, and shortened-$H$-band filters (0.9 - 1.7\um). This paper presents the design, performance, and early on-sky science of the WINTER observatory. 

\end{abstract}

\section{Introduction} \label{sec:intro}

Despite the recent rise in optical time-domain surveys, such as ATLAS \citep{atlas}, DeCAM \citep{DECAM}, Pan-STARRS \citep{pan}, ZTF \citep{ztf2019}, and GOTO \citep{goto} and the upcoming Rubin Telescope’s LSST survey \citep{Rubin}, the near-infrared (near-IR) time-domain sky remains relatively unexplored. High sky backgrounds and the cost of the state-of-the-art HgCdTe (Mercury Cadmium Telluride) sensors prevent similar growth in near-IR time-domain astronomy. Legacy and contemporary near-IR surveys---for example, TMSS \citep{tmss}, 2MASS \citep{2mass}, UKIRT \citep{UKIRT}, VISTA \citep{VISTA}, WISE/NEOWISE \citep{wise}, and Palomar Gattini IR \citep{De:2019xhw}---often trade off between coverage and depth, resulting in either relatively shallow surveys (for example, Palomar Gattini IR, $\text{J}\sim16$ mag; \cite{De:2019xhw}) or are restricted to small areas on sky (such as the  UKIRT Infrared Deep Sky Survey (UKIDSS) which covered  $\sim$7,500 deg$^2$ \citep{UKIRT} or the VISTA Variables in the Via Lactea (VVV) survey covering 520 deg$^2$ \citep{VISTA}). Consequently, in near-IR wavebands, there is a gap in deep, all-sky reference images and characterization of near-IR transients which would benefit the time-domain community and lay groundwork for the upcoming Nancy Grace Roman Space Telescope.

The Wide-Field Infrared Transient Explorer (WINTER) is a new near-IR time-domain survey instrument on a dedicated 1-meter robotic telescope at Palomar Observatory \citep{Lourie:2020, Frostig:2020}. WINTER covers over 1 deg$^2$ with a 90\% fill factor in the $Y$ (1.0 \um), $J$ (1.2 \um) and shortened $H$-bands (1.6 \um). The instrument was commissioned in June of 2023 and is operating robotically each night, surveying around 400 degrees per week to an average depth of $J_{\text{AB}} \sim 18.5$ magnitudes. 

WINTER is designed around three primary goals: (1) rapid robotic follow-up of gravitational-wave alerts to search for kilonovae, (2) near-IR time-domain surveys, and (3) general observing support for the WINTER community \citep{Frostig:2020}. The near-IR colors uniquely study the r-process materials produced in kilonovae, allowing two weeks or more to search for electromagnetic signatures from gravitational-wave events \citep{Frostig:2022}. Beyond kilonovae, near-IR observations are well suited for dust-obscured, cool, and intrinsically red transients, variables, and static sky science. Between following up gravitational-wave alerts, WINTER scans the accessible northern sky in a series of Galactic and extragalactic surveys studying a range of science  including  transiting exoplanets around low-mass stars, tidal disruption events, supernovae, stellar mergers, and young star outbursts (See Section \ref{sec:science} for a detailed introduction to WINTER science cases).  Over a ten-year nominal lifespan, WINTER will also build up a deep, co-added image of the near-IR sky, preparing for future infrared surveys and studying galactic structure, high-redshift quasi-stellar objects, and brown dwarfs.

In addition to WINTER’s science drivers, the instrument demonstrates the feasibility of Indium Gallium Arsenide (InGaAs) hybridized focal plane arrays as a cheaper alternative to HgCdTe sensors for near-IR astronomy. Unlike HgCdTe, InGaAs sensors do not require cryogenic cooling to achieve background-limited performance from the ground, further reducing the cost and complexity of the project. A prototype camera demonstrated this approach with InGaAs sensors and achieved sky-background limited photometry while running around -40$^{\circ}$C with a thermoelectric cooler (TEC) and secondary liquid cooling \citep{Simcoe:2019aps, Sullivan:2013}.  WINTER also features a novel ``fly's-eye optical" design to fill the 1 deg$^2$ field of view with six non-buttable sensors with only minimal gaps, and has custom readout electronics controlled by field programmable gate arrays (FPGAs), enabling fine control over the highly-programmable sensors. 


WINTER is part of a new class of InGaAs-based astronomical surveys. Parallel efforts in this category include the upcoming DREAMS telescope in the Southern Hemisphere \citep{DREAMS} and the smaller Antarctic Infrared Binocular Telescope (AIRBT; \citealt{Dong:2024, Dong:2025}). Together with the newly operating HgCdTe-based PRime-focus Infrared Microlensing Experiment (PRIME; \citealt{PRIME_performance, PRIME_camera, PRIME_optics}), these projects reflect a broader effort to expand coverage of the near-IR time-domain sky. These instruments further prepare for upcoming large-scale surveys such as the Nancy Grace Roman Space Telescope \citep{Spergel:2015} and Cryoscope \citep{Cryoscope}. 

This paper describes the WINTER instrument and early performance since its installation in June 2023. A more detailed treatment on WINTER's science cases can be found in Section \ref{sec:science}. Section \ref{sec:detectors} describes the InGaAs sensors and their performance, along with WINTER’s custom readout electronics, firmware, and software. The instrument design is described in Section \ref{sec:optics}, including the camera’s novel fly’s-eye optical design, the mechanical design, optomechanics, and cooling of the instrument. Section \ref{sec:obs} describes the observatory and 1-m robotic telescope, and Section \ref{sec:software} outlines the robotic control, scheduling, and data reduction software. Finally, the paper discusses on-sky performance, including a brief description of current surveys and preliminary results from select early science investigations (Section \ref{sec:performance}) and concludes (Section \ref{sec:conclusions}).

\begin{figure*}[t]
\epsscale{1.2}
\plotone{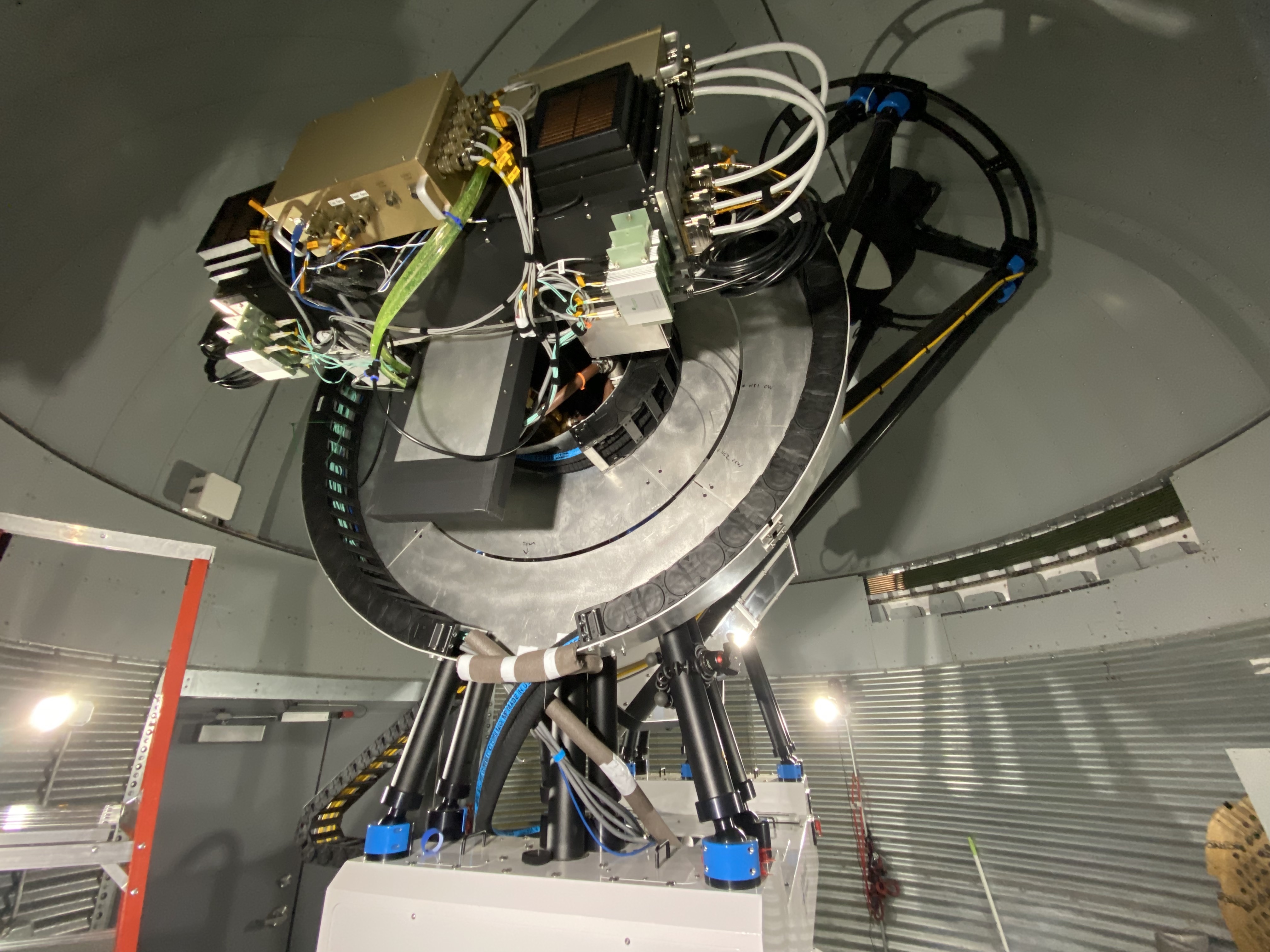}
\caption{WINTER mounted on its dedicated robotic telescope, including the enclosed optics and sensor box, electronics support boxes, liquid cooling, the filter tray and a cable wrap. 
\label{fig:winter}}
\end{figure*}

\section{Science Objectives}  \label{sec:science}
WINTER's design flows directly down from its science goals: rapid robotic follow up with low overheads, near-IR coverage, and a wide field of view \citep{Frostig:2020}. The science cases highlighted in this section introduce the motivation behind building the instrument, but are not a comprehensive list.

\subsection{r-process elemental synthesis} \label{sec:r_proc}

Near-IR observations are uniquely suited for studying kilonovae, thermal transients powered by rapid neutron-capture (r-process) nucleosynthesis in the neutron-rich ejecta from binary neutron star (BNS) or neutron star black hole (NSBH) mergers. The landmark BNS event, GW170817, detected in both gravitational waves and across the electromagnetic spectrum, demonstrated that r-process-rich ejecta obscure optical wavelengths, shifting the afterglow into the near-IR \citep{TheLIGOScientific:2017qsa, GBM:2017lvd, Kasen:2017sxr}. Unlike optical emissions, which are predicted to be short-lived and highly viewing-angle dependent, near-IR emission is both long-lived (over one week) and isotropic, making it a promising kilonova signature for BNS and some NSBH mergers \citep{Kasen:2013xka, Barnes:2016umi}. Studies show near-IR surveys could detect up to 8–10 times more kilonovae than optical surveys \citep{Zhu:2020ffa}.

To enable kilonova discovery, WINTER combines several key features: (1) sensitivity to the Lanthanide-rich kilonova emission peak in the near-IR; (2) a wide field of view; (3) fast robotic follow-up with low overhead resulting in the ability to quickly tile gravitational-wave uncertainty contours.

\cite{Frostig:2022} presents an end-to-end simulation of the WINTER follow-up observing of BNS mergers during O4. The study predicts that near-IR kilonovae are longer lived and red kilonovae are detected $\approx$1.5 times farther in the IR than in the optical. Detailed simulations of WINTER observing predicted up to five kilonova discoveries during O4. However, the sensitivities of both WINTER and the international network of gravitational wave detectors were lower than expected throughout O4. No BNS mergers were detected by LIGO, and despite several international follow-up campaigns of BBH and BH-NS mergers, there were no confirmed optical counterparts of GW events (e.g., \citealt{Pillas:2025}, including WINTER data).

\subsection{Peering through the dust}
Heavily obscured regions can conceal IR transients and variable sources. For example, the SPitzer InfraRed Intensive Transients Survey (SPIRITs) surveyed 200 nearby galaxies and identified 2457 variables and 78 transients over four years. Sixty of those transients were too red for optical detection, suggesting that optical surveys may miss up to $\sim$40\% of core-collapse supernovae in dusty environments \citep{Kasliwal_SPIRITs_2018, Jencson_2019}. SPIRITs also uncovered a new class of intermediate-luminosity transients, bridging the gap between novae and supernovae \citep{Kasliwal_SPIRITs_2017}.
IR monitoring also probes variability in young stellar objects (YSOs) across diverse environments. In the Orion Nebula and the Large Magellanic Cloud, \textit{Spitzer} observations showed that approximately 50\% of YSOs exhibit variability, enabling studies of jet activity, accretion processes, and circumstellar dust evolution \citep{yso_orion, yso_mag, yso_var}. IR surveys further trace dust-obscured Galactic structure. The VVV survey revealed previously unknown variable stars and star-forming regions in the Galactic plane, while UKIDSS extended sensitivity to fainter stellar populations in crowded and highly extincted fields \citep{VISTA, UKIRT}.

\subsection{Studying cool objects}

IR observations are essential for characterizing cool objects such as brown dwarfs, planets, red giants, and asymptotic giant branch (AGB) stars. These objects emit most of their flux at IR wavelengths due to their lower effective temperatures—below 300 K for brown dwarfs and up to 5000 K for red giants \citep{Faherty, Marocco}. Time-domain IR surveys reveal pulsation periods and episodic mass loss in red giants and AGB stars, tracing dust production and circumstellar evolution \citep{Liljegren}. Long-period variables (LPVs), including Miras and semi-regular AGB stars, show complex light curves that surveys such as WISE and VVV have systematically characterized \citep{Groenewegen, Nikzat}.\textit{Spitzer} monitored brown dwarfs with time-resolved spectroscopy and photometry, detecting rotational modulations caused by cloud structure and temperature variations. These observations inform models of substellar atmospheres \citep{Metchev, Artigau}. IR monitoring also tracks variability in cool-object binaries. Systems such as symbiotic stars produce IR variability through accretion, jet activity, and dusty outflows driven by interactions with compact companions \citep{Miko}.

Exoplanet transit surveys targeting M-dwarfs also benefit from IR observations. These cool, low-mass stars dominate the stellar population of the Milky Way and provide favorable conditions for detecting transits in red-optical and IR bands \citep{Dressing, Charbonneau}. The WINTER prototype InGaAs camera demonstrated this capability by observing exoplanet transits at Las Campanas Observatory \citep{Simcoe:2019aps}.

\subsection{Static sky science} \label{sec:static}

Over its ten-year lifespan, WINTER will build a deep reference image of the near-IR northern sky. The reference image will aid transient detection, support nightly source identification, and contribute to mapping L and T dwarfs, quasars, and Galactic structure. The static reference image and a long-baseline time-domain survey will provide valuable preparation for the upcoming time-domain surveys from the Vera C. Rubin Observatory and Nancy Grace Roman Space Telescope. 

\section{Sensors, readout electronics, firmware, and software} \label{sec:detectors}

\begin{figure*}[]
\epsscale{1.15}
\plotone{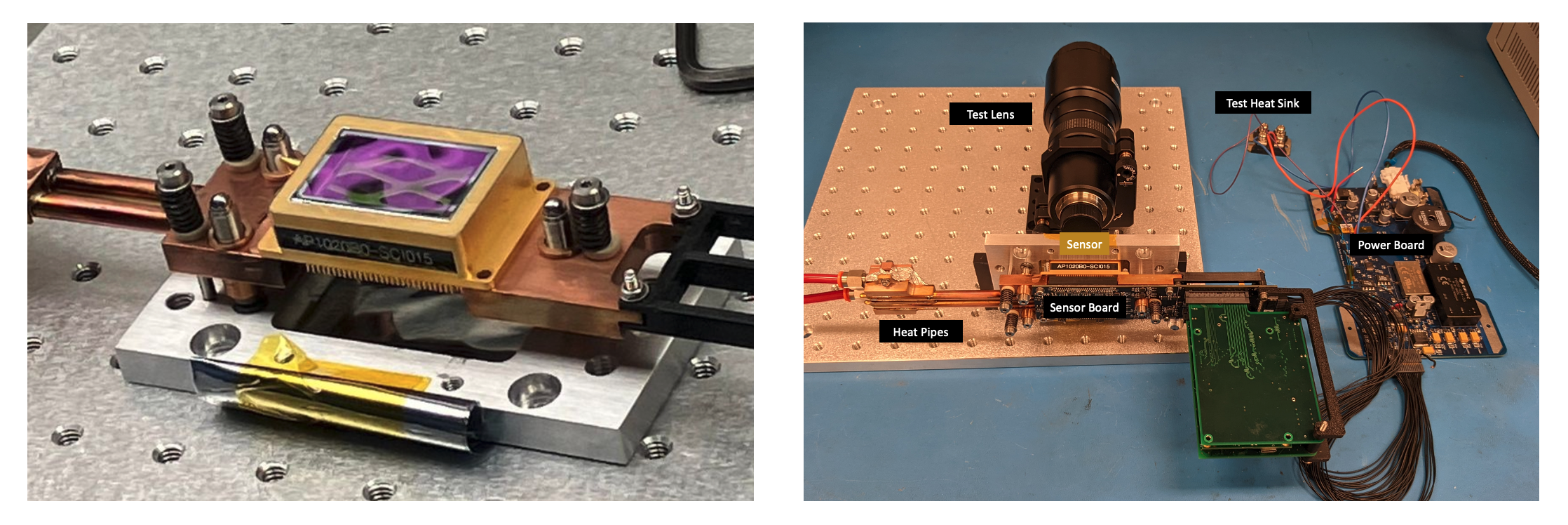}
\caption{One of the WINTER InGaAs sensors (left)  with custom readout electronics (right). The sensor is cooled by a two-stage TEC in the vacuum-sealed housing. The housing is mounted to a copper sled with copper heat pipes drawing heat away from the electronics to a heat exchanger leading to a liquid cooling loop. The five readout boards power the sensor and provide local control of each sensor through an Artix 7 FPGA. 
\label{fig:electronics}}
\end{figure*}

\subsection{InGaAs sensors}
The six InGaAs sensors central to WINTER are custom AP1020 focal plane arrays (FPAs) developed in collaboration with Teledyne-FLIR and Sensor Creations (now Attollo Engineering Inc./Safran Defense \& Space Inc.). With a high-definition format of 1920 by 1080 pixels and a 15\um pitch, the sensors are the largest format InGaAs sensors ever manufactured for commercial use (Figure \ref{fig:electronics}). InGaAs exhibits dark current 2--3 orders of magnitude lower than 1.7 \um cutoff HgCdTe at equivalent temperature \citep{Seshadri_2007_MCT_vs_ingaas_JPL_2007} down to $-100$~C, or in the  170--300~K range. For ground-based imaging applications like WINTER, InGaAs sensors achieve background-limited performance without the need for cryogenic cooling. Their 1.7 \um cutoff also reduces sensitivity to thermal background emission, unlike 2.5 \um HgCdTe detectors, which require cryogenically cooled optics.

The hybridized complementary metal oxide semiconductor (CMOS) focal plane arrays consist of photosensitive lattice-matched InGaAs PIN (p-type/insulator/n-type) diodes hybridized to a readout integrated circuit (ROIC) with a capacitive transimpedance amplifier (CTIA) pixel architecture. With a lower-energy bandgap than silicon, InGaAs is sensitive in the near/shortwave-infrared (NIR/SWIR) between 0.9 \um and its 1.7 \um room-temperature direct bandgap cutoff wavelength. The highly-programmable ROIC allows for multiple imaging modes, field-programmable gain, and application-specific sensor tuning. In contrast to the source follower per detector (SF) readout common in astronomy detectors like the Teledyne Hawaii family, the CTIA architecture supports video rate imaging at the price of increased read noise \citep{ROIC_pixel_architectures_guellec_2014}. An optimized CTIA readout can also deliver a higher degree of linearity than SF. To help mitigate this increased read noise (see Section \ref{sec:rn}), the AP1020 ROIC features options for on-chip correlated double sampling and multiple non-destructive reads which substantially reduce the effective read noise. The sensor read out with eight interleaved data output channels, allowing for full frame, continuous sampling at a frame rate of up to 30 Hz. The focal plane array is bonded to a two-stage thermoelectric cooler (TEC) in hermetically-sealed vacuum housing (see further details in Section \ref{sec:thermal}).

\subsection{Readout}
Custom electronics, firmware, and software are required to control the InGaAs sensors and to accommodate WINTER’s size, heat, and readout speed requirements. The sensors are individually managed by dedicated electronics controlled by FPGAs, sending data over fiber to a computer located in a nearby shed to minimize heat buildup in the dome. This configuration allows for low-noise and low-latency data acquisition within a relatively compact, freely rotating instrument. The custom readout scheme also leverages the fast readout of the sensor and works in tandem with robotic observing to enable near-real-time transient detection and alerts. Further discussion on the WINTER readout design can be found in \cite{Malonis:2020} and \cite{Frostig:2022SPIE}.  

\subsubsection{Electronics}
WINTER’s focal plane electronics are split into five boards, assembled into a focal plane module for each detector (Figure \ref{fig:electronics}): 
\begin{enumerate}
    \item The sensor board plugs directly into the InGaAs sensor and serves as a motherboard, connecting the other boards to the sensor with minimal signal path lengths. 
    \item The analog board converts the image data to digital signals with eight high-speed, 16-bit analog to digital converters (ADCs).
    \item The FPGA module is an Opal Kelly XEM7310 board with a Xilinx Artix 7 FPGA, controlling the other boards and the InGaAs sensors and providing image clocking, housekeeping, and power control. 
    \item The digital board interfaces between the FPGA module and the other electronics.
    \item The power board distributes power to the sensor and electronics and hosts a thermoelectric cooler  (TEC) drive circuit and a housekeeping system for closed-loop temperature control. 
\end{enumerate}

\subsubsection{Firmware and software}
The FPGA firmware runs the sensors, TEC, and focal plane readout electronics locally and interfaces with the WINTER control software. The firmware is written in Verilog, fully modeled in Vivado Simulator, and provides three main functions: electronic control, sensor control, and interfacing with the software. To operate the sensors, the firmware provides clocking, imaging signals, and writes the highly programmable ROIC settings, enabling imaging in various modes at a range of frame rates. In addition, the firmware controls power sequencing, telemetry, housekeeping, and data collection of the sensors and readout electronics, along with a safety watchdog that oversees system temperatures and connectivity to the software. Finally, the firmware communicates with the software through a USB 3.0 connection using the Opal Kelly Host Interface, providing an interface to enact commands from the software and to collect and stream sensor data. 

The six FPGA modules stream data over USB 3.0, converted to OM4 multimode fiber, to a readout computer in a neighboring shed. Each sensor connects to its own Icron USB 3.0 Spectra 3022 extender, which provides full-bandwidth data transfer, extends the USB range to a remote readout computer outside the telescope dome, and electrically isolates the link to prevent ground loops. An independent multi-threaded, {\tt\string python}-based software daemon independently controls each sensor in parallel for rapid data readout and temperature monitoring. A top level daemon coordinates the image acquisition on each sensor and collates the data into a single multi-extension \texttt{FITS} file.  

\subsubsection{Read-out modes}
The WINTER sensors feature two implemented read-out modes: integrate while read (IWR) and non-destructive read (NDR). IWR is a simpler read out mode where a single image is sampled at the end of the exposure. In this case, the data are read out while the subsequent image is exposing to reduce overheads. This simpler mode was deployed during commissioning of the WINTER sensors and is the current operating mode, since exposures are not read-noise limited. Alternatively, NDR allows for a reduction in read noise (see Section \ref{sec:rn}) and is an available mode for observing with WINTER, although it is not currently used in the survey. Below we describe the read-out scheme for NDR.

For short exposures, data are streamed in at seven frames per second, utilizing every available frame, and for longer exposures, the FPGA streams in every other frame, pausing between frames to control temperature deviations. At the end of an exposure, the continuously sampled frames are immediately pre-processed into single-frame, six-sensor FITS format images which are then downlinked from the observatory for further data reduction (see Section \ref{sec:drp}). To pre-process the WINTER data, the many readout frames are collapsed into one image with one of two techniques: (1) correlated-double sampling (CDS), in which the final frame is subtracted from the first frame, or (2) sample-up-the-ramp (SUTR), in which a linear fit is used to get a slope of the signal in each pixel \citep{NDR}. SUTR reduces read noise and recovers data affected by saturation, cosmic rays, and streaming errors \citep{Robberto_2014}, but also adds significant data volumes and overhead time. 

To facilitate transient discovery, WINTER requires rapid data pre-processing and reduction to minimize data storage volume and enable real-time decisions on candidate events and observing strategies. The use of graphics processing units (GPUs) is well-suited to efficiently process high-definition imaging data. WINTER’s readout computer has two commercial NVIDIA GeForce RTX 3080 Ti GPUs, each one processing three sensors’ images in parallel. For ease of use, the GPU code uses {\tt\string cupy}, an open-source GPU-implemented {\tt\string python} library modeled after {\tt\string numpy}. Implementing the nonlinearity correction and NDR linear fit on GPUs leads to a factor of $\sim$10 speed up from a CPU-only implementation for a five minute exposure with six sensors. 

\subsection{Thermal control}
\label{sec:thermal}
WINTER’s InGaAs sensors use a vacuum-sealed package with a two-stage thermoelectric cooler (TEC) at the core of their cooling system. Each sensor has its own mechanical and thermal module which includes a kinematic mounting system for holding the detector. The mount allows for manual focus adjustments during assembly and provides the required cooling to the detectors' TEC modules. An oxygen free high conductivity (OFHC) copper sled, sandwiched between the back of the sensor and the readout electronics, draws heat away from the sensor with two sintered copper/water heat pipes leading to a liquid heat exchanger, placing any liquid outside of the main instrument box and away from the optics and electronics \citep{Frostig:2022SPIE}. The heat pipes are soldered with solder paste into a OFHC copper flange which is bolted to a small liquid heat exchanger with an indium foil interface material to maximize conductive coupling. An OptiTemp OTC 1.0A closed-loop 70\%/30\% distilled water/propylene glycol mix chiller located in a shed next to the telescope dome provides parallel cooling to each of the six focal plane modules through a manifold located on the telescope rotator. The chiller is set to 10$^{\circ}$C which maintains the warm side of the TEC to 12$^{\circ}$C through the $\sim$15 m of hose between the chiller and the instrument. The sensor readout computer controls the TEC setpoint with a proportional-integral-derivative (PID) loop. The instrument itself is sealed with a series of gaskets to keep out dust, pollen, and other environmental contaminants common at Palomar Observatory. To avoid condensation within the camera enclosure during humid days, a dry air purge system consisting of a 1 horsepower Powerex oil-less compressor and a nano purification systems dryer delivers air with a -40$^{\circ}$C dewpoint. This system delivers 28 liters per minute of air which is piped through the camera with a series of 1/4 inch hoses with small holes drilled in them, which provides sufficient airflow through the box to keep critical electronics like the FPGA modules from overheating.

\subsection{Performance}

The WINTER project commissioned the first batch of these InGaAs sensors, which included designing a new, larger ROIC. The vendor manufactured the sensors on a best-effort basis and enough sensors were produced to fill in WINTER's focal plane; however, the sensors did not meet several key performance specifications, as discussed in this section. In total, fourteen sensors were received, with varying levels of functionality. Six of the highest-performing sensors were integrated into WINTER, with five non-responsive units allocated for thermal testing and three kept as spares. Overall, this deviation from expected performance results in a factor of $\sim$10 reduction in instrument efficiency ($\sim$2 magnitudes decrease in depths) from the instrument design.

\begin{table}[b!]
\centering
\caption{A comparison of the predicted instrument performance and the measured performance for a few key metrics. The measured dark current and read noise assume a conversion gain of g=2.5 e$^-$/DN.}
\label{tab:discrepancy}
\begin{tabular}{|l|l|l|l|}
\hline
\textbf{Metric}    & \textbf{Predicted}                         & \textbf{Measured} & \textbf{Units} \\ \hline
Gain & \parbox[t]{2.0cm}{\raggedright 1 (spec.)\\1.6 (design)} & 2.5 & e$^-$/DN \\ \hline
\parbox[t]{1.5cm}{\raggedright Quantum efficiency} & 0.80                                       & 0.05--0.10        &                \\ \hline
Dark current       & 113                                        & 120--370          & e$^-$/s/pix    \\ \hline
Read noise         & 45                                         & 50--70            & e$^-$          \\ \hline
\end{tabular}
\end{table}

\begin{figure}[t]
\epsscale{1.0}
\plotone{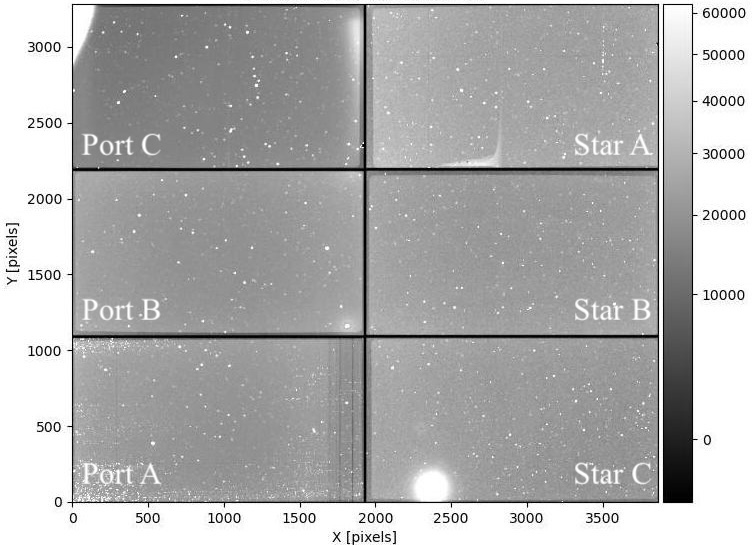}
\caption{A raw image of WINTER's focal plane with six sensors. The sensors are named by position in the physical instrument, including a port and starboard side with three sensors each. The panel presents the six sensor frames arranged to mirror their layout on the focal plane, with the color scale indicating the measured counts. Prominent artifacts include areas where sensors have partially failed due to glow-spot contamination—seen in the Port C panel and the circular imprint in the Star C—as well as variations in dark current driven by thermal shifts, evidenced by the higher background levels in the Port B relative to Port C above it.
\label{fig:raw}}
\end{figure}

\subsubsection{Conversion gain} \label{sec:gain}

\begin{figure}[ht!]
    \centering
    \epsscale{1.1}
    \plotone{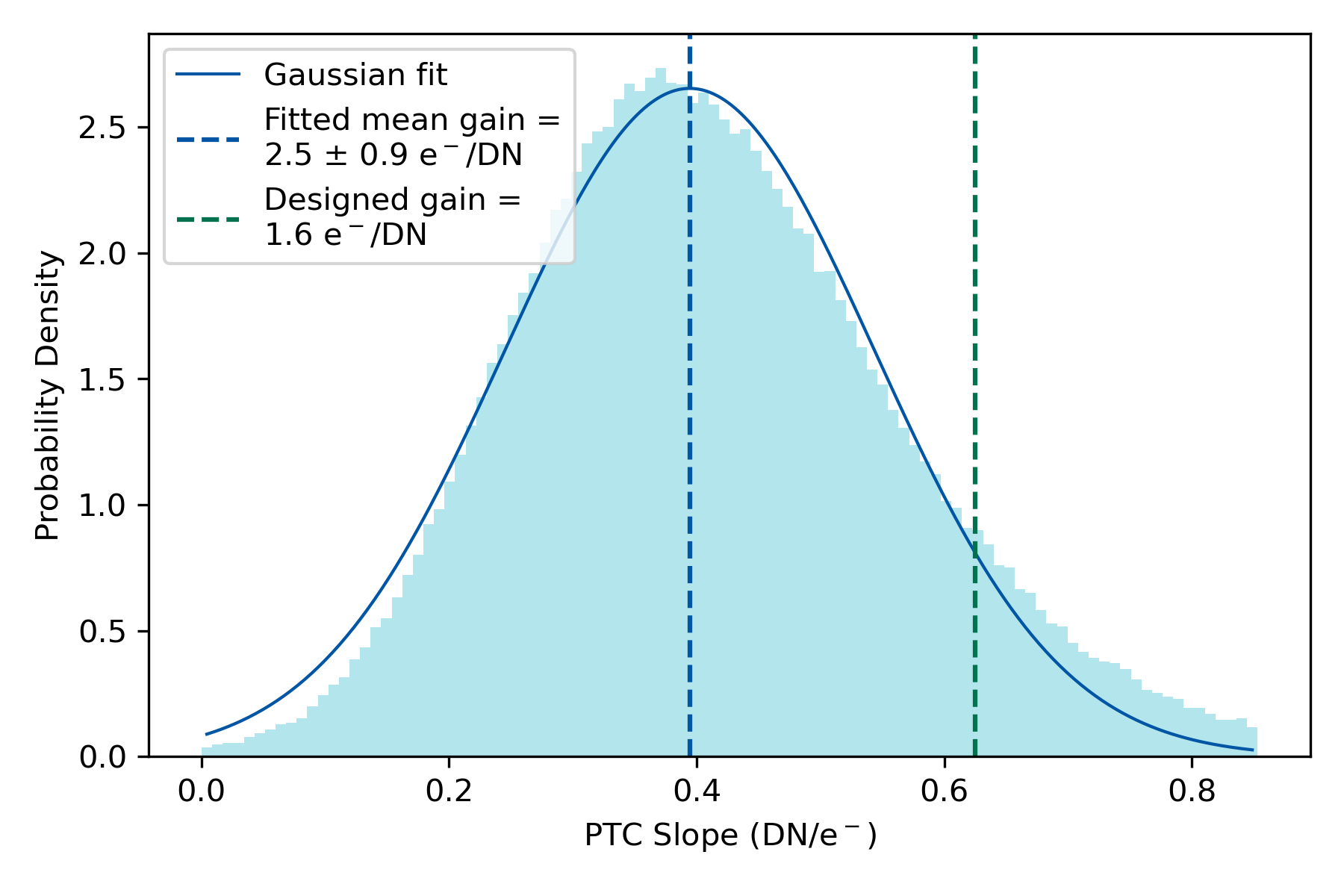}
    \caption{A histogram of photon transfer curve (PTC) slopes measured for individual pixels within one readout channel of a WINTER sensor. The PTC slope (DN/e$^{-}$) is determined from the linear relationship between variance and mean signal in flat field images at different exposure levels. The inverse of this slope yields the conversion gain for each pixel. The distribution shows 255,000 individual pixel measurements (light blue histogram) with a Gaussian fit overlaid (dark blue line). The fitted mean gain is 2.5 $\pm$ 0.9 e$^{-}$/DN (1$\sigma$ standard deviation of the pixel-to-pixel variation). The designed gain specification of 1.6 e$^{-}$/DN (green dashed line) falls within the distribution but is significantly lower than the fitted mean.}
    \label{fig:ptc}
\end{figure}

Characterizing the WINTER sensors begins with measuring the conversion gain that links photoelectron signal (e$^-$) to the corresponding digital output (in digital numbers, DN). Based on the ROIC design, the gain is estimated to be g=1.6 e$^-$/DN (see \cite{Frostig:thesis} for the full calculation). Typically, this value is experimentally verified using a Photon Transfer Curve (PTC) measurement, which assumes the presence of a shot-noise limited regime where noise increases with signal level $\text{N}$ as $\sqrt{\text{N}}$. Multiple PTC approaches explored in \cite{Frostig:thesis}, reveal nonlinear responses (see Section \ref{sec:linearity}) and differing gain estimates for the NDR and IWR read out modes. Each method produces a gain estimate within a $60\%$ error of the expected g=1.6 e$^-$/DN. Furthermore, the gains vary by sensor, between the eight read out channels within each sensor, and across the pixels within those read out channels. 

For simplicity, we present one measurement of the gain here and use it throughout the paper, but note the significant spread in gain by pixel, requiring careful flat-fielding. Figure \ref{fig:ptc} shows the range of slopes from a PTC for one readout channel of a WINTER sensor, taken with a set of laboratory flats on an integrating sphere. The variance is taken for a stack of ten dark-subtracted frames at each exposure time, with eight rounds of sigma clipping to reject pixels outside of three standard deviations of the mean. A line is fit to the slope of the variance versus mean signal for each pixel. This analysis predicts a mean gain of g=2.5 e$^-$/DN but shows a spread of responses across pixels within this read out, with some pixels consistent with the designed gain of g=1.6 e$^-$/DN. Given this uncertainty, we present all analyses throughout this study in both measured counts (DN) and converted to electrons (e$^-$).

We measured a quantum efficiency (QE) of less than 10\% for the WINTER sensors, significantly lower than the designed 80\%, indicating substantial losses in photon-to-electron conversion efficiency. This measurement was validated through three independent tests: (1) laboratory characterization of the full system using a controlled light source, (2) direct laboratory measurement of the sensors’ photocurrent response, and (3) on-sky calibration with reference stars. Laboratory measurements were made using a 20-inch integrating sphere illuminated with a laser driven light source from Energetiq and a series of optical and near infrared narrow bandpass filters. The absolute optical flux was determined using a silicon photodiode with a NIST-traceable responsivity calibration, allowing the quantum efficiency to be measured between 500 and 1050 nm. 
The test measured a peak response of approximately 5\% QE, suggesting a manufacturing issue leading to deviations from the expected efficiency (Figure \ref{fig:qe}) Furthermore, in standard QE and photon transfer curve (PTC) measurements, conversion gain and QE are inherently degenerate, necessitating an independent test to pinpoint the root cause of the discrepancy.

\subsubsection{Quantum efficiency} \label{sec:qe}
\begin{figure}[]
    \centering
    \epsscale{1}
    \plotone{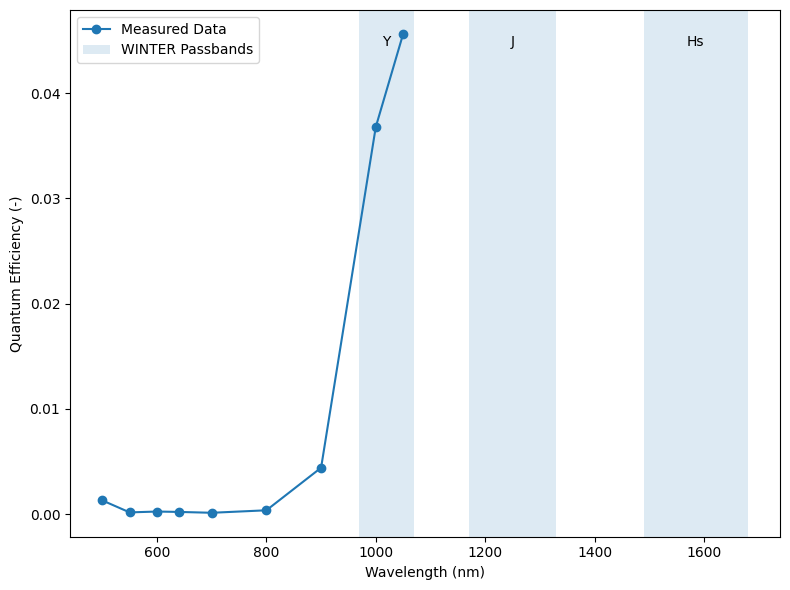}
    \caption{Measured quantum efficiency (QE) of one WINTER sensor at visible and near-IR wavelengths. A low response is expected outside of the near-IR (WINTER's three filter wavebands are highlighted in blue); however, the measured near-IR QE is lower than expected at $\sim$5\% as opposed to the designed 80\%. This test assumes a gain of 2.5 e-/DN.}
    \label{fig:qe_lab}
\end{figure}

\begin{figure*}[ht!]
    \centering
    \plotone{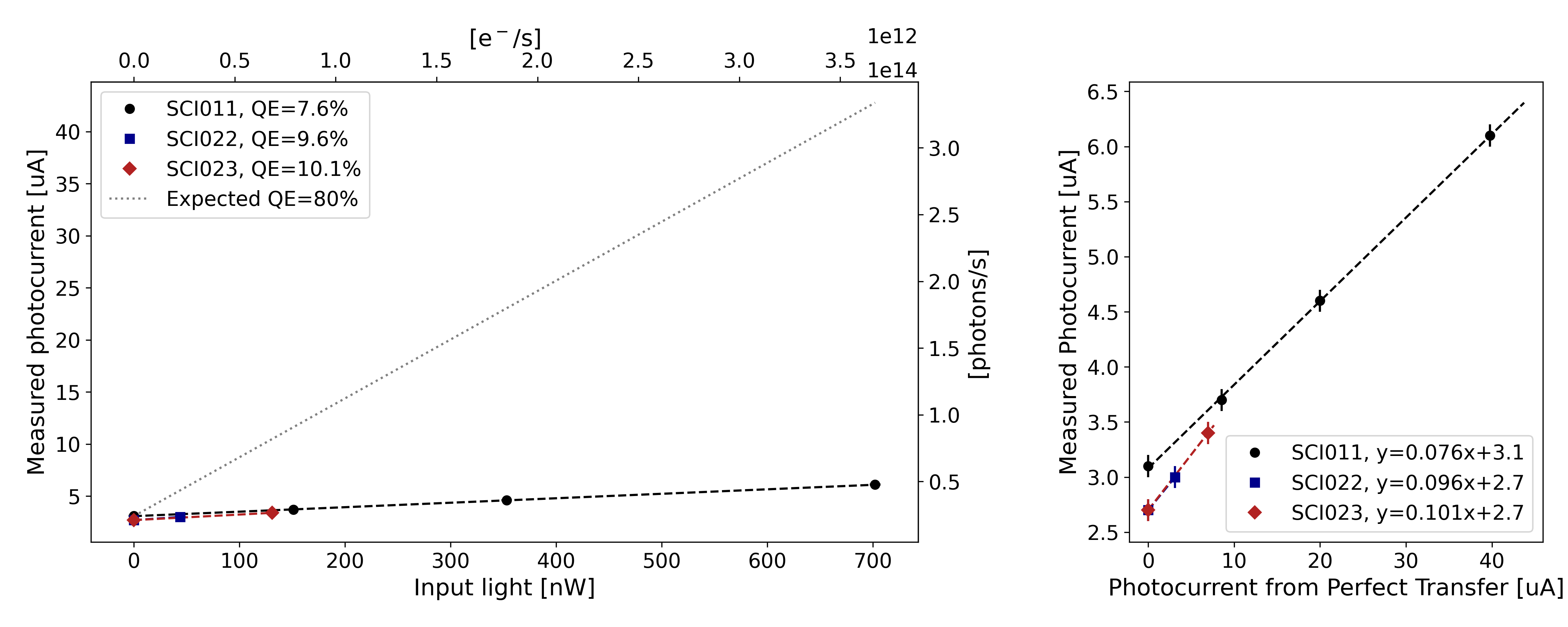}
    \epsscale{\textwidth}
    \caption{Direct testing of the WINTER sensor quantum efficiency (QE) by measuring the effect of external light on the sensor bias voltage, bypassing any effects of the read-out electronics or sensor control logic. Left: the measured photocurrent (also shown in photons/s) for an input lamp brightness (also shown in e$^-$/s), for three sensors used in testing compared with the expected QE of 80\%. Right: Measured photocurrent compared to the predicted photocurrent under perfect charge transfer, $I_{\mathrm{perfect}}$, derived from the known input optical power and photon energy. The slope determines the effective QE, showing a range of 8\%-10\%, compared to the specified 80\%. \\ }
    \label{fig:qe}
\end{figure*}

To isolate the sensor photodiode array's intrinsic response to light from any effects introduced by the ROIC or the custom readout chain, we conducted a secondary set of QE tests. These tests employed an external source measurement unit to directly supply the sensor bias voltage, allowing precise voltage control and direct photocurrent measurement on the sensor while bypassing the readout circuitry via programmable logic on the ROIC. This approach enabled a direct measurement of photocurrent response, confirming the measurement of a peak QE of $10\%$ (Figure \ref{fig:qe}). We define the quantum efficiency as the ratio of the number of electrons generated per incident photon. Given an input optical power \( P \) (in nW) and a measured photocurrent \( I \) (in µA), the QE can be expressed as:

\begin{equation}
    \text{QE} = \frac{I \cdot h \cdot c}{P \cdot e \cdot \lambda}
\end{equation}

\noindent where \( I \) is the measured photocurrent in amperes (A), \( P \) is the incident optical power in watts (W), \( h  \) is Planck’s constant (J·s), \( c \) is the speed of light (m/s), \( e \) is the elementary charge (C), and \( \lambda \) is the wavelength of the incident light (m). 

To further evaluate the sensor’s response, we consider the relationship between the measured photocurrent and the expected photocurrent from perfect charge transfer (Figure \ref{fig:qe}). The expected photocurrent under ideal conditions, \( I_{\text{perfect}} \), is compared to the actual measured photocurrent \( I_{\text{meas}} \). The effective quantum efficiency can be determined from the linear regression slope in the relation:

\begin{equation}
    I_{\text{meas}} = \alpha I_{\text{perfect}} + I_0
\end{equation}

\noindent where \( \alpha \) represents the QE, and \( I_0 \) accounts for any systematic offset in the measurement. Figure \ref{fig:qe} shows the results of this testing for three spare WINTER sensors, one of which (SCI022) was on-sky for one year. The three sensors show QEs between 6\% and 10\%, significantly lower than the specified design of $>80$\% in all bands. 

Finally, on-sky flux observations of standard stars independently measure WINTER’s full instrument and telescope throughput. Accounting for as-measured reflectivities of the mirrors and lenses, the results confirm an average sensor QE of $\sim$10$\%$ (see \cite{Frostig:2024} for a detailed calculation). This analysis includes sensor serial number SCI022, which was previously in the Starboard B position in the full instrument, which has a laboratory measured QE of 9.6\% (Figure \ref{fig:qe}). 

Given the results of the three independent tests, the decreases QE likely arises from a manufacturing or quality control issue during hybridization, photodiode array production, or subsequent processing. The test procedures confirm the low QE is not due to deficiencies in the ROIC or amplification/read-out chain. Since the InGaAs growth and photodiode array design made use of mature processes at Teledyne-FLIR, we conclude that this particular production lot likely encountered problems in the preparation or bonding of the InGaAs layer to the silicon multiplexer—a hypothesis that could be verified through destructive testing and microscopy of a sensor from the batch.

\begin{figure}[th!]
\epsscale{1.2}
\plotone{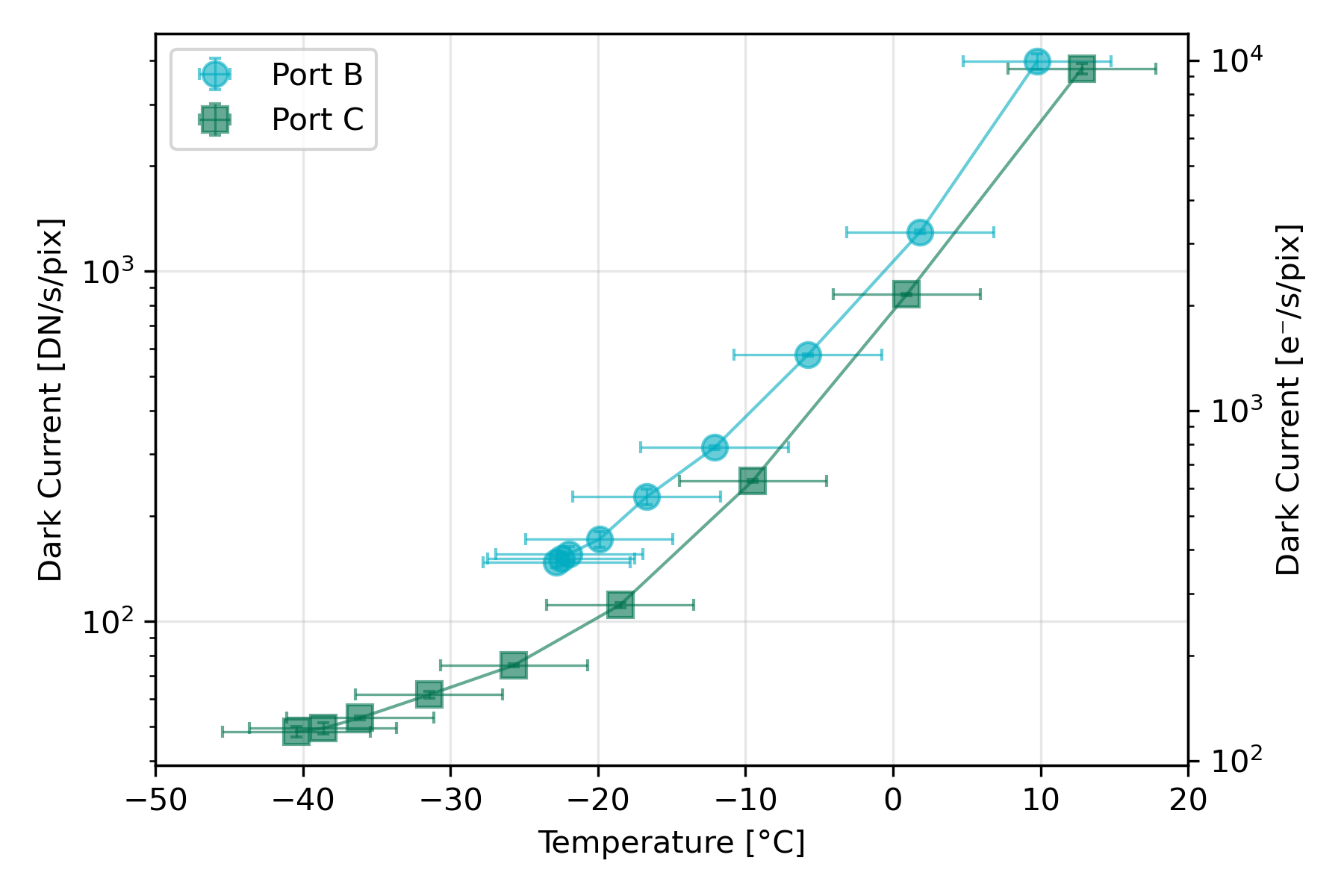}
\caption{Dark current as a function of sensor temperature for the warmest (Port B) and coldest (Port C) WINTER sensors, measured on sky. The sensors do not reach the designed operating temperature of T $=-50^{\circ}$C, due to under sizing of the TEC and parasitic heat paths in the package design. There is also significant variation in coldest operating temperature between sensors ($-25^{\circ}$ to $-40^{\circ}$C). The dark current follows an expected linear decrease with log(Temperature) until about $-15^{\circ}$C, where the response levels out, potentially due to thermal infrared emission from the detector housing. Values are plotted both in DN/\,s/\,pix and converted to e$^{-}$/\,s/\,pix using a gain of 2.5\,e$^{-}$/\,DN, corresponding to a lowest dark current of 120 and 370 e$^{-}$/\,s/\,pix, for Port B and Port C, respectively. There is an uncertainty in the temperature calibration of $\pm$5$^{\circ}$C.
\label{fig:dark}}
\end{figure}
 
\begin{figure}[t!]
\epsscale{1.2}
\plotone{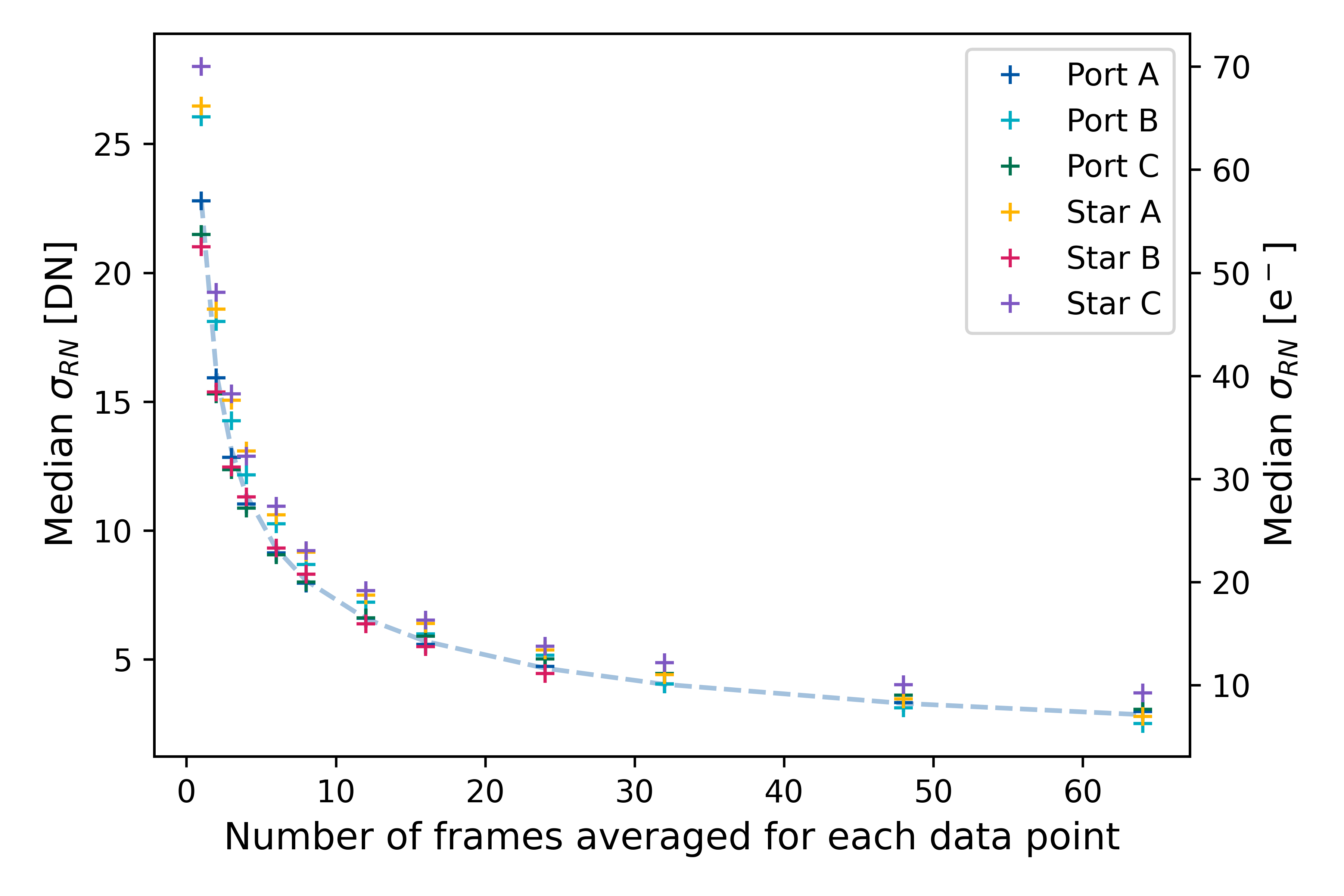}
\caption{Read noise (in measured counts and electrons) in the six on-sky WINTER sensors, demonstrating the reductions in read noise achieved by using the nondestructive read (NDR) mode. In the nominal integrate while read (IWR) mode, the image is only read once, leading to read noise between 50 and 70 e$^-$ RMS. In NDR, multiple reads of the image are taken while integrating, allowing for techniques to reduce the effective read noise per final image. The example shown demonstrates how using 64 frames in the read out reduces the read noise to $\sim$5 e$^-$ RMS. For the Port A sensor, a dashed line demonstrates the theoretically achievable reduction in read noise, decreasing with a factor of $\frac{1}{\sqrt{N}}$ for N averaged frames. Data are shown in measured counts (DN) and converted to e$^-$, assuming a gain of 2.5 e$^-$/DN for all pixels.
\label{fig:rn}}
\end{figure}

\subsubsection{Dark current} \label{sec:dc}

WINTER is designed for sky-background-limited photometry, with TEC-based cooling at T $=-50^{\circ}$ C to reduce dark current levels below the sky background. The smaller prototype instrument employed a single-stage TEC and demonstrated background-limited performance between -$40^{\circ}$C to -$45^{\circ}$C, with dark current halving for every 7$^{\circ}$C of cooling \citep{Simcoe:2019aps}. However, for the manufactured WINTER sensors, a combination of TEC under-scope and insufficient process control in sensor packaging resulted in decreased cooling capacity and varied final operating temperatures across the six sensors. 

Figure \ref{fig:dark} displays the dark current in the six WINTER sensors by temperature in the final integrated camera on the telescope. The sensors reach a minimum temperature between T $=-20^{\circ}$ C and T $=-40^{\circ}$ C, unable to achieve the designed cooling of T $=-50^{\circ}$ C. The temperature discrepancy arises from two key factors. First, the TEC selection process did not include realistic modeling of the sensor packaging, resulting in the use of an undersized two-stage TEC (Laird MS2-107-10-10-12-12-11-W8). Second, the final design placed an excessive heat load on the TEC through parasitic radiative coupling and thermal conduction through wire bonds from the pins. Temperature variations between sensors are due to differences in TEC bonding, vacuum seal integrity, and the temperature of each sensor board, which impacts the TEC’s hot-side temperature. The dark current decreases approximately linearly with the logarithm of the temperature down to about $-15^{\circ}$C, beyond which it plateaus, possibly due to thermal infrared emission from the detector housing.

Figure~\ref{fig:dark} also highlights the sensor-to-sensor variation in dark current among WINTER’s six on-sky sensors. Part of the apparent spread reflects the $\pm 5^{\circ}$C uncertainty in the on-chip temperature calibration, but intrinsic differences are also present. Port C—consistently the coldest device—exhibits roughly a factor of six lower dark current than Port B at their respective operating temperatures. These relative background levels are visually apparent in the focal-plane image shown in Figure~\ref{fig:raw}.

\subsubsection{Read noise}
\label{sec:rn}
To evaluate the read noise in the WINTER sensors we analyze the reference rows of each sensor. These reference rows, positioned outside the active 1920 × 1080 photosensitive region, are insensitive to incident radiation. As a result, any signal variations in these pixels arise predominantly from read noise without incorporating the effects of pixel response nonlinearity or photon shot noise. Figure \ref{fig:rn} illustrates an analysis of the signal standard deviation for each reference pixel over 384 frames in a non-destructive read ramp. The median read noise for a single readout, equivalent to the current on-sky IWR mode, is approximately 24.5 DN RMS (equivalent to 61.3 e$^-$, assuming a gain of 2.5 e$^-$/DN), which is close to the initial project requirement of 45 e$^-$ RMS, depending on the exact gain per pixel.

By continuously reading out the sensors with non-destructive sampling, multiple exposures of the same pixel can be averaged before a reset, reducing the random electronic noise in accordance with the theoretical $\frac{1}{\sqrt{N}}$ scaling. Figure \ref{fig:rn} demonstrates how this technique lowers the effective read noise as frames are binned in increasing groups from two to 64. The standard deviation of the readout signal along a best-fit line decreases, with the lowest read noise reaching approximately 3 DN (7.5 e$^-$) when averaging in groups of 64 frames. 

Despite achieving read noise levels within project specifications, even with a single read, a low-frequency read-noise component introduces a time-dependent striped pattern in the images. This structured noise is effectively removed in the data reduction pipeline using Fourier filtering (see Section \ref{sec:drp}).

\subsubsection{Linearity} \label{sec:linearity}
\begin{figure}[]
\epsscale{1.2}
\plotone{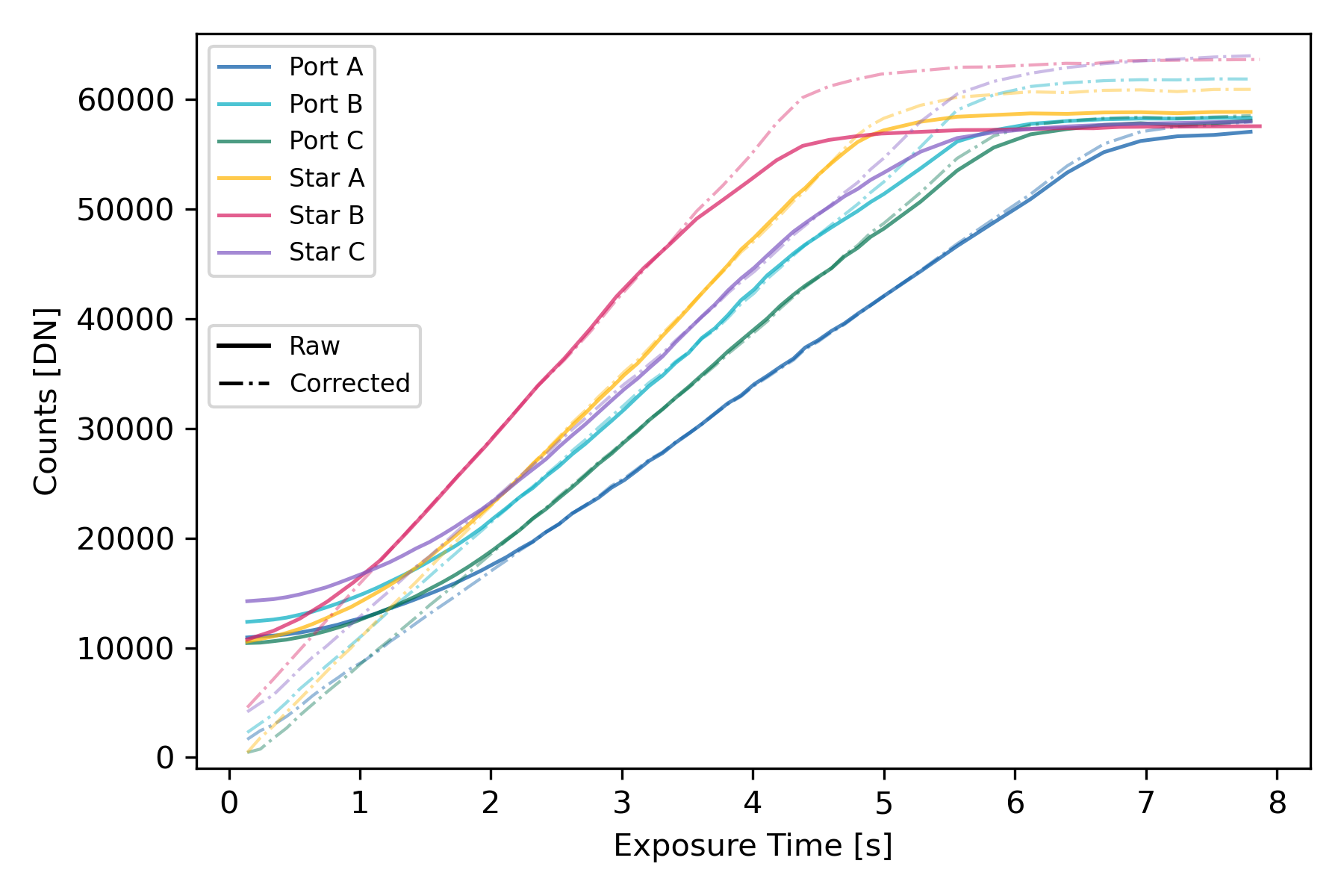}
\caption{Nonlinearity seen in the WINTER sensors, showing the response to uniform illumination for a central, example pixel on the six on-sky WINTER sensors (solid lines). The response shows significant low-end nonlinearity. All images are corrected with a pixel-wise nonlinearity correction, which uses a two-part polynomial fit to to model the pixel response and linearize the response (dashed lines). 
\label{fig:nonlin}}
\end{figure}

WINTER’s sensors exhibit stronger nonlinearity than originally expected, introducing additional challenges for photometry of faint sources (Figure \ref{fig:nonlin}). The most pronounced deviations occur at low to moderate signal levels, with measurable curvature beginning near 10,000 counts and continuing well into the mid-well regime. Even in the region of highest linearity, roughly 30,000–45,000 counts, the response is not fully linear.

To correct for these effects, we derive a per-pixel piecewise polynomial calibration. Each pixel is modeled with an 11th-order polynomial for the low-count region and a 5th-order polynomial for the high-count region, joined at a fixed crossover point of 25,000 counts with a 10,000-count overlap region. The fitting uses uniformly sampled exposure-time sequences of dome flats. During science processing, raw counts are passed through the appropriate segment of the piecewise model to yield an approximately linearized response, as seen in the dashed lines of Figure \ref{fig:nonlin}. The nonlinearity correction results in a median nonlinearity $\textless0.5$\% across the WINTER sensors. The effect of these corrections on the science data is explored in V. Karambelkar and R. D. Stein et al., in prep. 

To make a bad-pixel mask, a line is fit to the central 33–85\% of each pixel’s dynamic range—the portion of the well that behaves most linearly—and residuals relative to this baseline are evaluated across the full well. Pixels whose residual standard deviations exceed 500 ADU, including unresponsive, unstable, and hot pixels, are flagged and added to the bad-pixel mask.

The nonlinearity correction coefficients, bad pixel masks, generation, and implementation code are all released publicly through the \texttt{winternlc} package available on git.\footnote{https://github.com/winter-telescope/winternlc}

\section{Instrument Design} \label{sec:optics}
\begin{figure*}[ht!]
\plotone{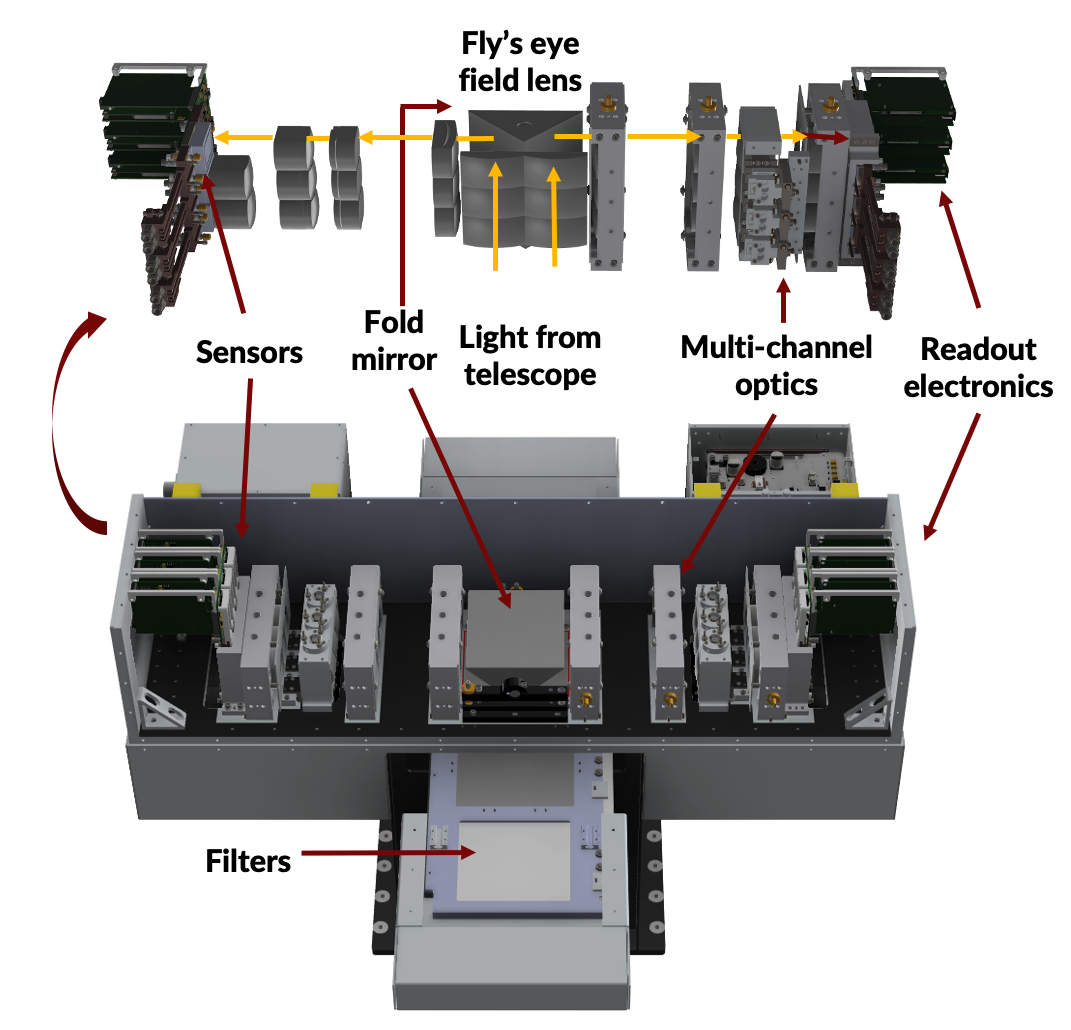}
\caption{A rendering of the WINTER optical and optomechanical design. Below: a model of all of the instrument components in its housing, with the back and top panels removed, shown from the back. Above: details of the design shown from the front, with one isolated sensor assembly and the bare optics (upper left) along with just the optomechanical alignment structures (upper right). The fly’s-eye layout splits the F/6 telescope beam into six parallel optical channels, each reimaged at F/3 onto its own InGaAs sensor, achieving a 1'' pixel scale matched to Palomar seeing. The figure also illustrates the truncated, bonded lens groups that allow close packing of the optics, the right-angle fold mirror that divides the channels between two sensor banks, and the custom mechanical mounts that align and support the optics while minimizing mass and stray light.
\label{fig:optics}
}
\end{figure*}
\subsection{Optical fly's-eye design}

WINTER’s optical design covers a 1 deg$^2$ field of view with $\textgreater$90\% fill factor despite the sensors’ vacuum-sealed housing preventing close tiling (Figure \ref{fig:electronics}). The novel fly’s-eye optical design splits the image between six identical optical channels to direct each section of the image onto its respective sensor with minimal inter-chip gaps between channels (Figure \ref{fig:optics}). To enable close packing of the optics, each lens is truncated symmetrically about one plane, matching the aspect ratio of the HD-format sensors. After passing through a waveband filter, the telescope’s focal plane is divided into six channels by a field-lens array made of six bonded lenses. A right-angle fold mirror then divides the image into two halves, placing three sensors on either end of the instrument to keep the center of gravity close to the telescope rotator \citep{Hinrichsen:2020}.

Furthermore, WINTER’s optical design prioritizes a 1\arcsec~pixel scale for the 15 \um sensor pitch to match the median seeing at Palomar Observatory and other time-domain surveys such as PTF and ZTF. This necessitates a 2:1 demagnification of the F/6 telescope beam to an F/3 beam at the sensor. Each optical channel consists of ten spherical lenses. Lenses 1-5 collimate the telescope F/6 beam onto a pupil between lenses 5 and 6, and then lenses 6-10 reimage the pupil at an F/3 beam onto its respective InGaAs sensor.

All lens groups are bonded into groups of three tiled lens elements, with the exception of the field lens array (lens 1), which has six bonded lenses, and the lens 6-7 doublet, which retains independent adjustment to act as a compensator lens for fine focus optimization. Tolerancing studies simulating alignment procedures for the WINTER optical system for using Zemax’s Monte Carlo simulations found that maintaining strict fabrication tolerances for the lenses allows for loosening of mechanical alignment tolerances \citep{Lourie:2020, Frostig:2020}. This approach simplifies the optical alignment procedure, allowing for vertically tiled lens elements to be aligned together with the lens 6-7 bonded doublet allowing for independent adjustment for each optical channel. All WINTER lenses were purchased from Optimax Systems, who co-developed the precision truncation and alignment of the bonded, tiled lens elements, and all glasses were purchased from Ohara Corporation from the same melt batch to reduce variation between optical channels. To verify infrared performance, an outside company, M$^3$, measured the index of refraction of a sample of each glass from 0.7 -- 1.7 \um. The infrared properties of these glasses had not yet been reported in the literature and closely matched predicted performance.

\subsection{Filters}\label{sec:filters}
\begin{figure}[t!]
\epsscale{1.2}
\plotone{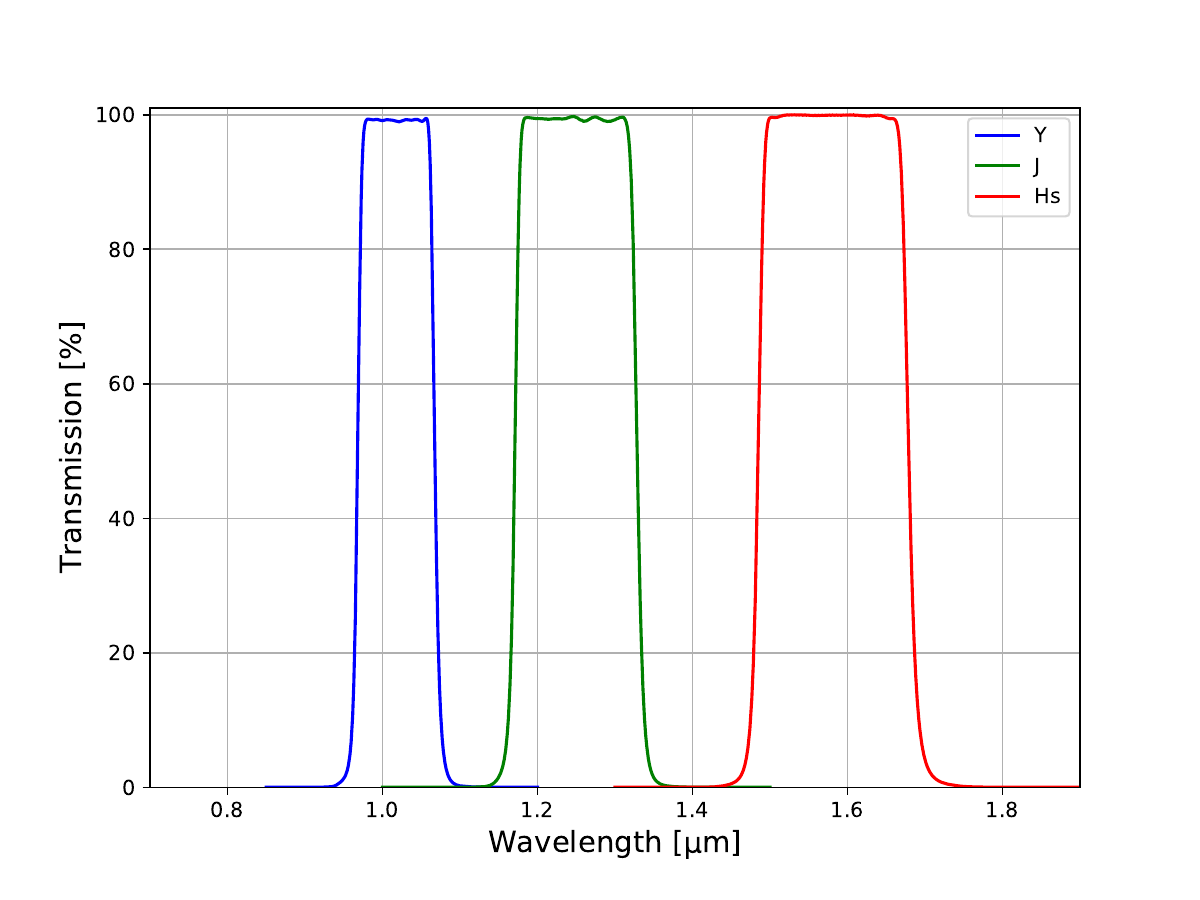}
\caption{The measured filter transmission of WINTER's $Y$, $J$, and $Hs$ filters, maufactured by Asahi Spectra on 10 mm fused silica (adapted from \cite{Lourie:2020}). 
\label{fig:filters}}
\end{figure}

WINTER observes with $Y$, $J$, $Hs$ (shortened H) filters based on the Mauna Kea Observatory filter set with a long-wave cutoff in H-band at 1.7 \um to match the InGaAs bandgap cutoff. The filters, manufactured by Asahi Spectra on 10 mm fused silica, achieve $\textgreater$99$\%$ in-band throughput while blocking out-of-band light to $\textless$0.01$\%$ (Figure \ref{fig:filters}). The filters are placed 50 mm in front of the instrument field lens array. Placement within the telecentric, F/6 converging beam of the telescope ensures spectral uniformity across the field of view, simplifying photometric calibration, with a modest reduction in the slope of the waveband edges due to the $\pm4.8^{\circ}$ angle of incidence variation within the F/6 beam.

\subsection{Filter tray} 
WINTER's aluminum filter tray, which holds three filters ($Y$, $J$, and $H_{S}$) and a mirror for dark frames, is mounted to the front of the instrument, perpendicular to the optical axis (Figure \ref{fig:optics}). A stepper motor with a pinion is coupled to a gear rack and actuates each filter change. The position of each of the filters is marked by a detent, which in conjunction with a spring-loaded arm, ensures the stability and stationarity of the filter tray when in use. Additionally, a linear encoder attached to the tray monitors the filter tray position to correct for any slippage that may occur during operation. Upon startup each night, the filter tray is homed to the dark mirror and each subsequent filter change between adjacent filters takes $\sim$15 seconds.

\subsection{Optomechanics} \label{sec:optomech}

\begin{figure*}[]
\epsscale{1.1}
\plotone{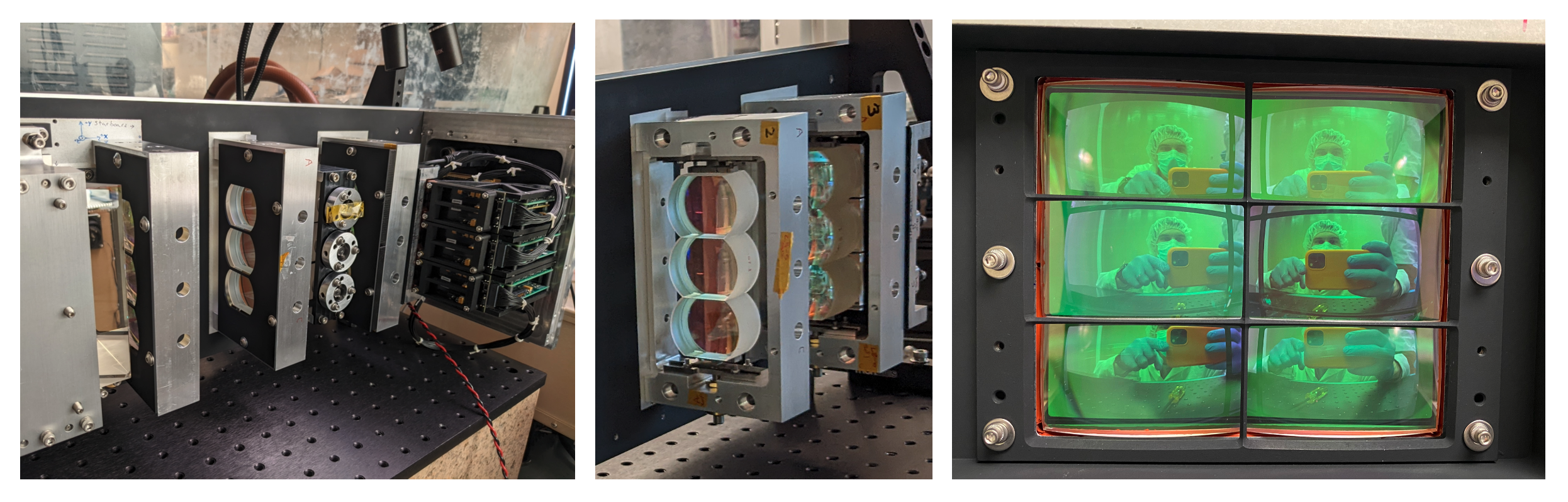}
\caption{An image of some select details of the optomechanical design and the stray-light mitigation baffles. Left: Half of the instrument during laboratory alignment, including baffled tiled lens groups, the compensator lenses, and the sensors and readout electronics. Middle: A tiled lens group mounted to the optical bench without a baffle. Right: The focal plane mask in front of the WINTER field lens to block stray light from inter-channel cross-talk. 
\label{fig:optics_details}}
\end{figure*}

The WINTER optomechanical design employs adhesive bonding for the lenses to reduce mass, design complexity, and stress on the optics. 3M Scotch-Weld 2216 A/B Gray epoxy was selected for all adhesive bonds based on extensive thermal and humidity lifetime tests simulating WINTER’s ten-year nominal lifetime \citep{Hinrichsen:2020}. The vertically-tiled lens elements are mounted in MIC-6 aluminum frames with flexure mounts made of Titanium 6Al-4V, providing single-axis lateral adjustment of the lens group (Figure \ref{fig:optics_details}). A custom 5-axis adjustable mount, made from Titanium 6Al-4V and Aluminum 7075-T6, holds the compensator lenses (6-7 doublet) with a stacked stage design where each stage adjusts one axis. The adjustment stages are accessible outside of the main instrument enclosure to provide flexibility for future focus modifications.  

The right-angle fold mirror is the heaviest optic in the instrument and the original adhesive bond design failed during initial alignment. Instead, the fold mirror has a six-point kinematic mount that fully constrains all degrees of freedom. The mount is loaded against six hardpoints with spring plungers, four on the front optical surfaces, one on the back face, and one on the top. To reduce the contact stresses, the spring plungers and hard points are all spherical contact surfaces with a machined Delrin cap to spread the load. All optomechanics are rated for a 6g shipping load, although the fold mirror was removed from the optical assembly during shipping.  

To mitigate stray light, the optomechanical design also includes a focal plane mask and baffles around the tiled lens groups (Figure \ref{fig:optics_details}). The focal plane mask is placed in front of the field lens to prevent cross-talk between optical channels and reflections off of the bonded edges of the tiled lenses, which was identified as a major source of stray light in the non-sequential ray-tracing simulations outlined in \cite{Frostig:2020}. The focal plane mask is milled from MIC-6 cast aluminum and painted with Krylon Ultra-Flat Black, with the openings matching the footprints of the sensors on the telescope focal plane. Similar aluminum baffles painted with Aeroglaze Z306 flat black are mounted to the tiled lens group frames, with openings oversized relative to the image footprint. Prototype focal plane masks made with selective laser sintering (SLS) 3D printers were produced as well. The 3D printed versions captured the complex geometry of the baffle but did not achieve the knife-edge sharpness of the machined masks. While these were not optically characterized, with further optimization this approach may be viable for future efforts.

\subsection{Alignment, integration, and test}
WINTER’s custom truncated lenses require special alignment, integration, and test (AIT) procedures. In this section, we provide a concise overview of the AIT procedure, while a comprehensive analysis of the instrument AIT process and outcomes will appear in an upcoming publication. First, the as-manufactured lens thicknesses were measured with a TriOpics OptiSurf interferometer, prompting a re-optimization of the lens spacings based on the tolerancing procedures outlined in \cite{Lourie:2020, Frostig:2020}. Next, the fold mirror was installed in the center of the instrument with a Hexagon coordinate measuring machines (CMM) arm used to verify the placement and tilt of the optic to a $\sim$25 \um accuracy. Small mirrors were attached to the sensor mounts on either end of the instrument and aligned with a Zygo Verifire wavefront interferometer to pre-align the tip, tilt, and focus position of the sensors before delivery from the vendor. The tiled lens group were then aligned with the CMM arm, with the aluminum frames resting against reference pins and shimmed to compensate for tip and tilt along the optical axis. The individually-mounted compensator lenses were then aligned with their five-axis mounts and verified with the CMM arm. For the final alignment, a fiber-fed pinhole mask manufactured by FiberTech Optica provided an array of five F/3 sources per sensor, allowing for fine optical three-axis positioning of the sensors and verifying the required instrument image quality. Due to time constraints and the initial alignment meeting design requirements, a final optical alignment with compensator lenses was not conducted.

\subsection{Camera housing/mechanics} \label{sec:mech}

\begin{figure*}[]
\epsscale{1}
\plotone{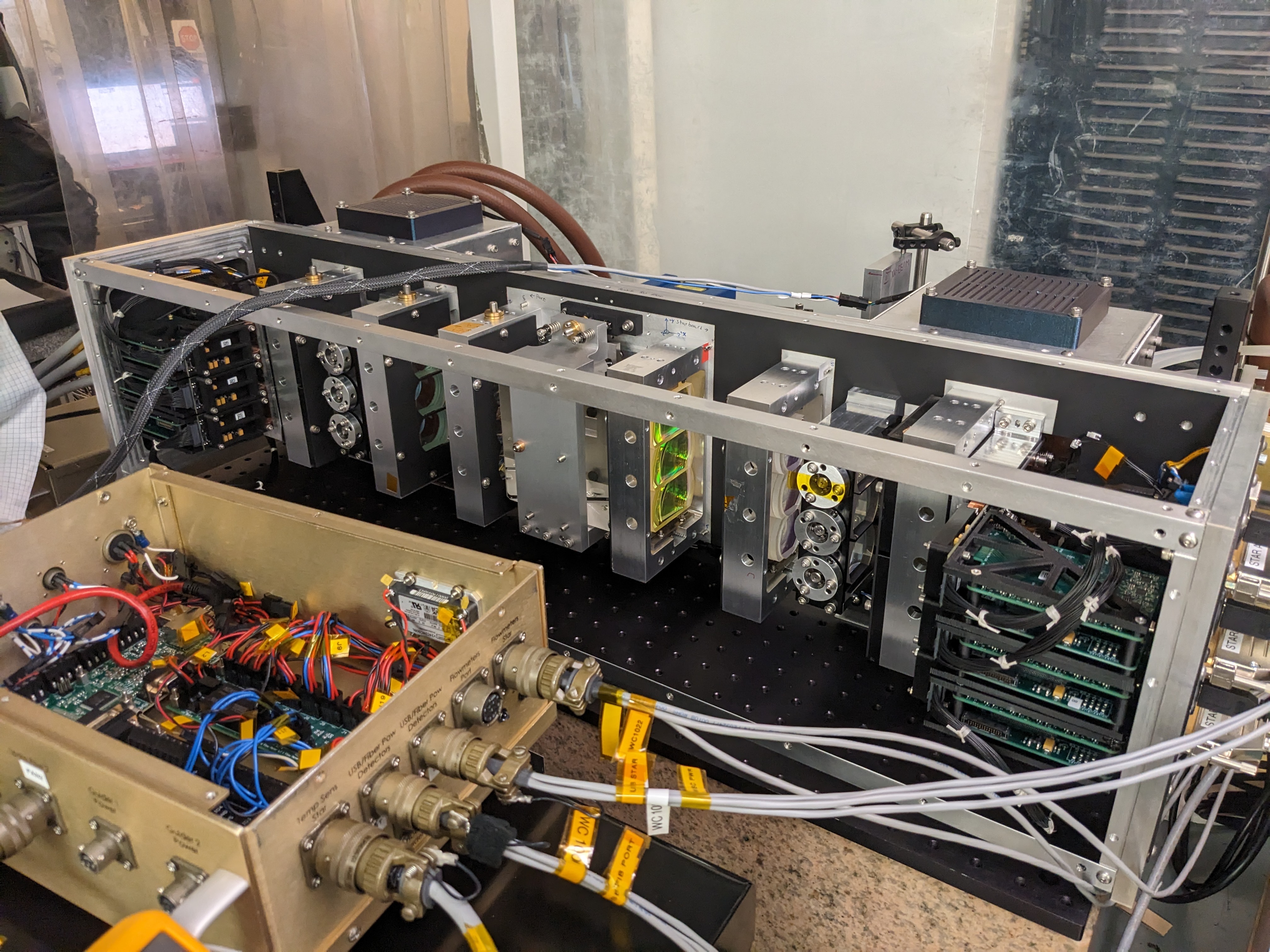}
\caption{The full WINTER instrument without the protective panels, showing the fly's-eye optics, custom optomechanics, sensor read out system, and power support. 
\label{fig:instrument}}
\end{figure*}

A precision optical bench serves as the reference plane for aligning and mounting all optomechanical components, including the sensor mounting frames (Figure \ref{fig:instrument}). The optical bench is made of a cast aluminum plate with flatness and parallelism tolerances of 0.025 mm and placement tolerance for mounting holes and alignment pins of $\pm$0.18 mm. It is painted with Aeroglaze Z306 flat black paint between non-critical surfaces to reduce stray light. Black-painted panels, resting on an aluminum frame and sealed with rubber gaskets to prevent dust contamination, form the remaining five sides of the enclosure. A focal plane mask located in front of the field lens array mitigates stray light due to cross-talk between optical channels (Figure \ref{fig:optics_details}). Similar baffles around the tiled lens arrays act as stray-light reduction and a safety retaining system in case of adhesive bond failure (Figure \ref{fig:optics_details}). To further prevent stray light near the image planes, Acktar Advanced Coatings’ Metal Velvet Black Foil was applied to the sensor mounting frames and parts of the filter tray.  A custom carbon fiber–reinforced plastic adapter, manufactured by CarbonVision, mounts the instrument to the telescope rotator.

\section{Observatory} \label{sec:obs}
\begin{figure}[]
\epsscale{1.2}
\plotone{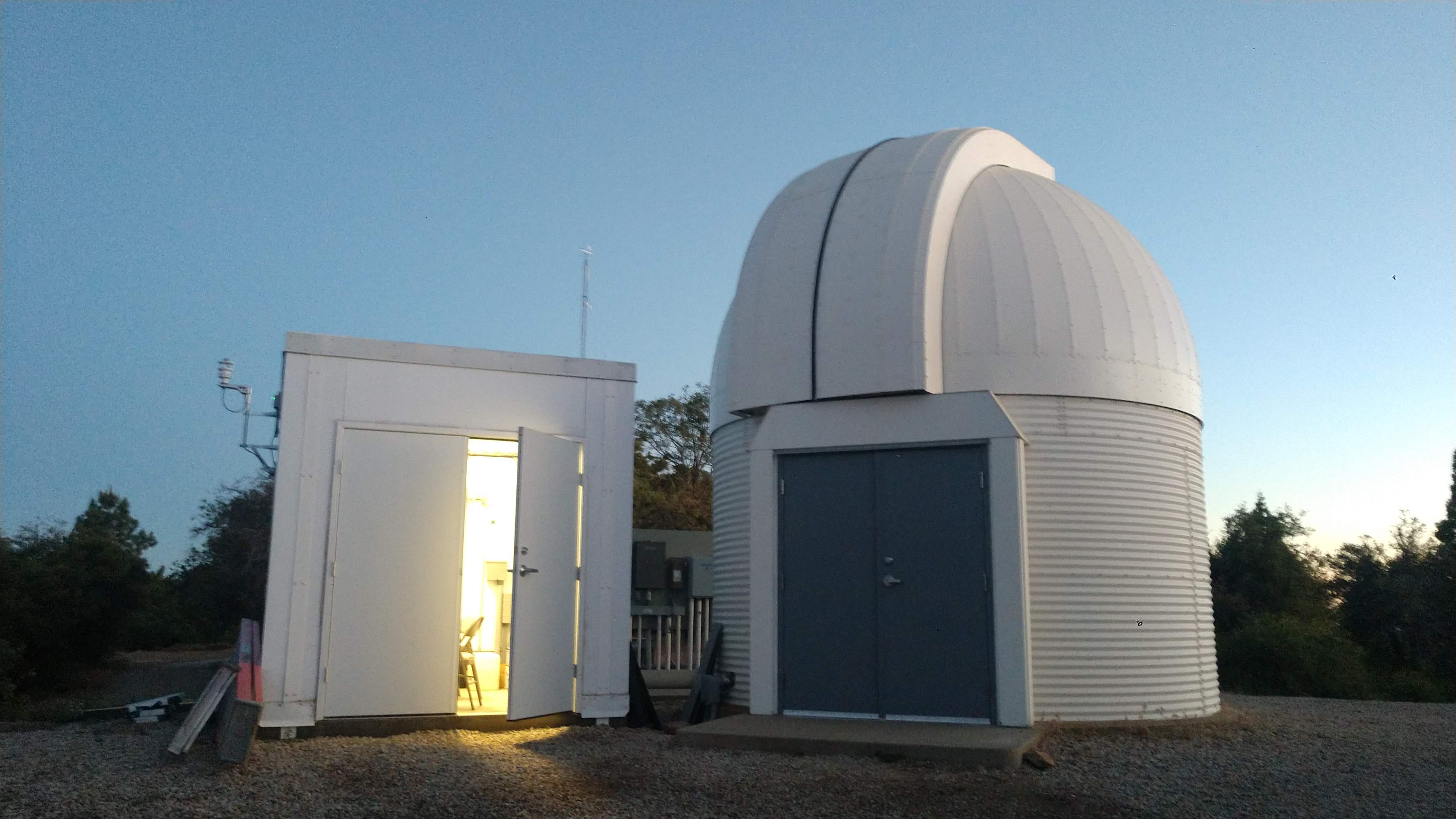}
\caption{The WINTER dome (right) and shed (left) at Palomar Observatory. The telescope, optics, and FPGAs are in the dome, while the shed houses the computers and chiller to keep sources of heat away from the instrument.
\label{fig:obs}}
\end{figure}

WINTER operates at Palomar Observatory in Southern California, sharing a site with the 200-inch Hale Telescope, the Zwicky Transient Facility (ZTF) \citep{ztf2019} and Palomar Gattini IR \citep{De:2019xhw}. The site is at an altitude of $\sim$1700 meters and is run by Caltech Optical Observatories. WINTER takes advantage of existing infrastructure at the site to run robotic telescopes, including data downlink, weather monitoring, and facilities operations. The weather and observing conditions are continuously monitored by telescope operators and support astronomers at the 200-inch Hale Telescope, who enable and lockout observing throughout the night accordingly. This continuous staffing support also allows for prompt hands-on response to operational anomalies and equipment failures that may occur over years of operation. 

The instrument operates in a preexisting five-meter dome with a new, adjacent climate-controlled shed housing the WINTER control computers and electronics, power supplies, and chiller (Figure \ref{fig:obs}). The dome was retrofit for robotic operations, including a robotic control and telemetry system similar to those developed for ZTF and PGIR, a new weather station, motor drives, and a custom telescope pier. The dome’s robotic control system allows for extensive autonomous safety protocols to protect both workers and equipment, as outlined in \cite{Frostig:2020}. 

\subsection{Telescope}

\begin{figure}[]
\epsscale{1.0}
\plotone{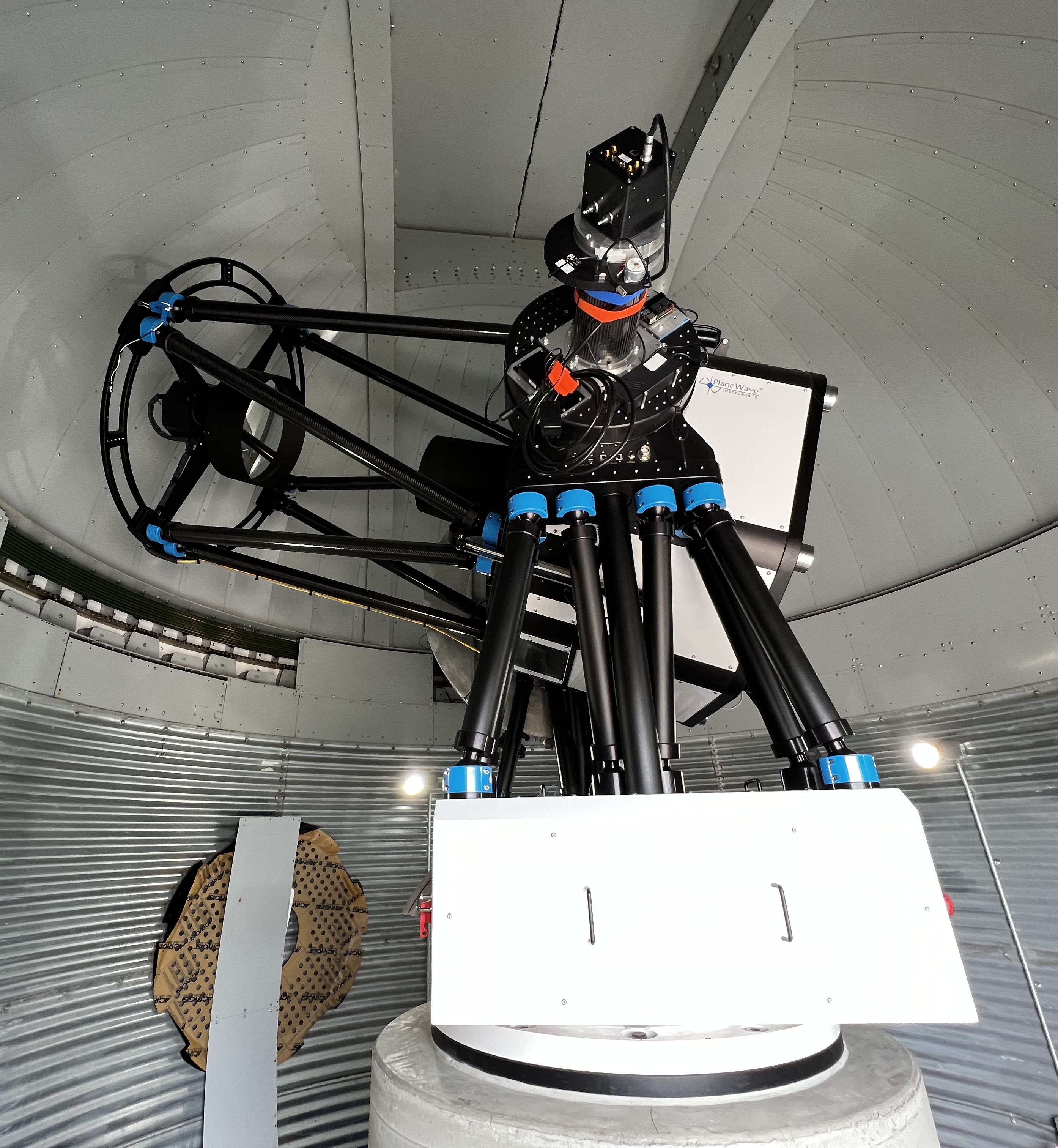}
\caption{The PlaneWave 1-meter robotic telescope with SUMMER on the companion Nasmyth port. WINTER is on the opposing port as seen in Figure \ref{fig:winter}.
\label{fig:obs}}
\end{figure}

WINTER employs a 1-meter commercial off-the-shelf (COTS) corrected Dall-Kirkham (CDK) telescope made by PlaneWave Instruments with modifications for the project (Figure \ref{fig:obs}). The PW1000 1-Meter Observatory System features an instrument rotator mount to correct for sky rotation and fast direct-drive altitude and azimuth motors with a maximum velocity of 15 degrees per second with a settle time between slews of one second. The telescope optics include a 1-meter primary mirror, a spherical secondary mirror, a flat tertiary mirror with an integrated rotator to select between the two Nasmyth ports, and a set of three corrector lenses at each of the ports. All of the telescope optics are made of fused silica to increase thermal stability. Customizations for the WINTER project include developing a focusing system at the secondary mirror instead of the default focus mechanism at the instrument rotator mount, and applying near-IR-optimized anti-reflection coatings on the corrector lenses for one Nasmyth port. The opposing port retains optical-wavelength-optimized coatings and houses a companion instrument.

\subsection{SUMMER and SPRING}
The telescope and a companion instrument, SUMMER (Studying the Universe with Multi-Messenger and Exoplanet Research), were installed in June of 2021, ahead of WINTER’s completion in 2023, enabling development of the robotic control code, scheduler, and data reduction pipeline (see sections \ref{sec:robo_control_inst}, \ref{sec:scheduler}, \ref{sec:drp}) and performing optical follow up of science targets (e.g., \cite{Christos:2023}). The instrument consisted of a Raptor Photonics CCD with e2v 42-40 architecture, which includes a 5-stage TEC and liquid cooling loop, a 65 mm optical shutter from Vincent Associates, and a Finger Lakes Instrumentation seven-position filter wheel which includes u, g, r, and i SDSS passband filters from Asahi Spectra. SUMMER images cover a 0.29-degree-squared field of view. 

Since commissioning the telescope, the second port on the 1-meter telescope has also served as a testbed for new sensors. There have been multiple observing runs with optical Complementary Metal-Oxide-Semiconductor (CMOS) sensors, including the Hamamatsu Photonics qCMOS camera and the large-format Teledyne COSMOS camera (see results in \citealt{Layden:2025}). Currently, it is home to the SPRING (Super PalomaR INGaas) camera, a Princeton Infrared Technologies 1280SciCam InGaAs camera with 1280 x 1024 12.5 \um pixels used for deep infrared observations of targets that fit within the smaller 8.8 x 7.0 arcminute field of view.

\begin{figure*}[]
\epsscale{0.95}
\plotone{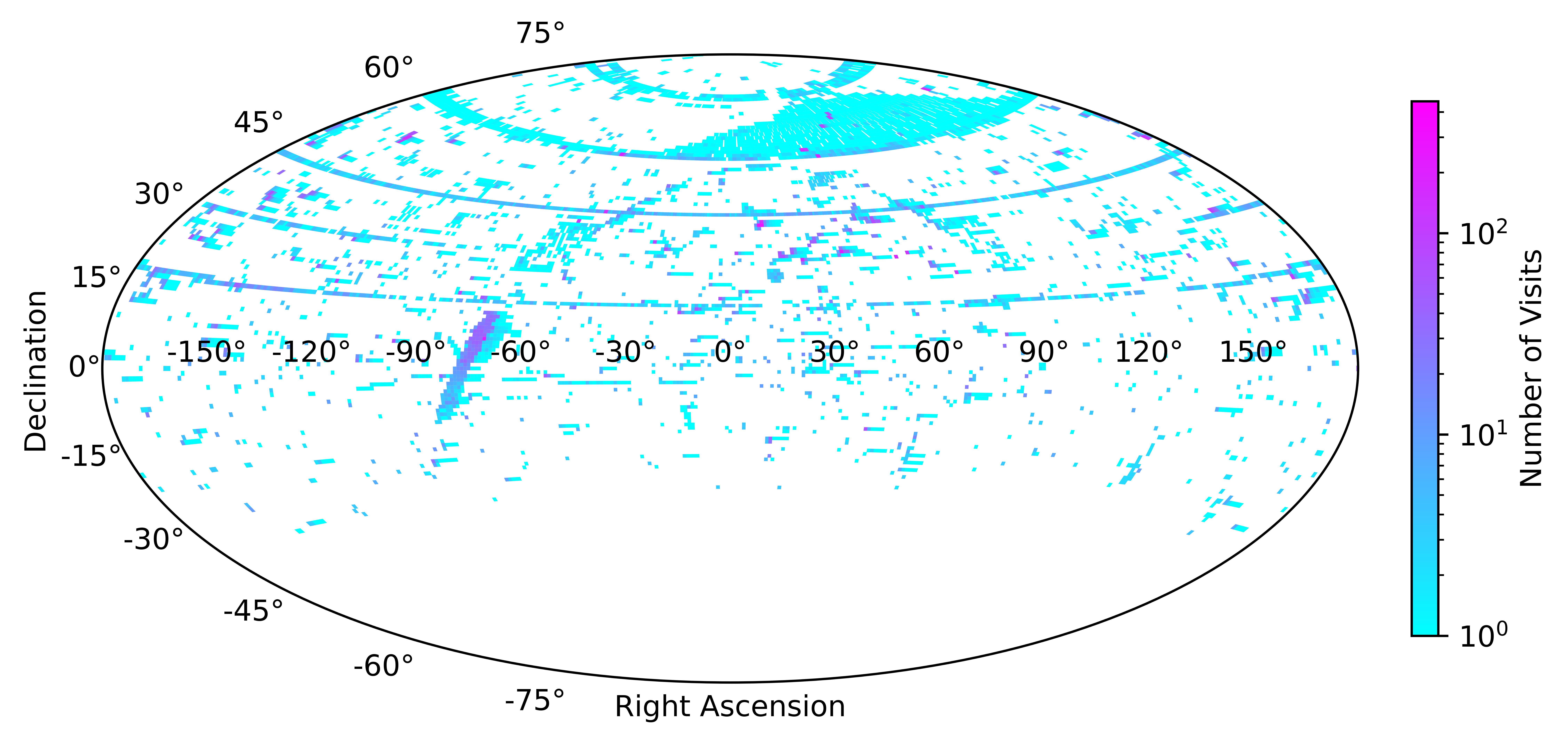}
\caption{A cumulative map of all WINTER observations between June 2023 and December 2025. Surveys so far include the galactic plane (a seen in a central arc), nearby galaxies, and building up reference images (as seen in stripes of constant declination).
\label{fig:scheduler}}
\end{figure*}

\section{Software} \label{sec:software}
\subsection{Robotic control software} \label{sec:robo_control_inst}
The WINTER telescope's operations are managed by the WINTER Supervisory Program (\texttt{wsp}), a highly parallelized Python 3.9-based control system leveraging PyQt5\footnote{https://riverbankcomputing.com/software/pyqt/} for interfacing with the Qt application framework\footnote{https://www.qt.io/development/qt-framework}, enabling sophisticated control and scheduling capabilities \citep{wsp_software_spie} \footnote{https://github.com/winter-telescope/observatory/}. The WSP consists of more than a dozen individual subsystem daemons which communicate with each other over TCP/IP sockets using the \texttt{Pyro5} python library\footnote{pyro5.readthedocs.io}, each of which are multi-threaded to allow concurrent commanding, communication, and housekeeping monitoring. The system is launched and managed as a Linux system daemon on the observatory's control PC, with individual subsystem daemons run on separate PCs. \texttt{wsp} generates nightly schedules and manages observation tasks. This system, drawing upon methodologies from the MINERVA \citep{MINERVA_Swift_2015}, Robo-AO \citep{robo-ao_Riddle_2012}, GOTO \citep{goto_dyer} and SuperBIT \citep{superbit_romualdez} projects, facilitates automated decision-making for observatory operations, such as dome opening and observation initiation, while allowing manual override by operators in case of adverse conditions. \texttt{wsp} orchestrates a variety of processes including scheduling, Target of Opportunity (ToO) event handling, command execution via a priority queue system, and comprehensive logging of telescope states and observations through real time telemetry graphing and \texttt{slack} notifications. 

\subsection{Scheduler and ToOs} \label{sec:scheduler}

To autonomously balance multiple observing programs with a range of cadences, filters, and priorities, WINTER employs a custom scheduling software based on the ZTF scheduler \citep{ztf_scheduler_Bellm_2019}. It creates nightly schedules based on WINTER’s surveys and long-term targets, which can be interrupted by ToO requests at any point in the night. A flexible API, \texttt{winterapi}\footnote{https://github.com/winter-telescope/winterapi} allows authorized users to submit ToO requests directly to the queue, download data associated with their observing programs, and enables high-level tracking by the team to monitor observing time. This framework allows observers to manage their own programs, and enables automated follow-up triggering of transient candidates without the need for an observer in the loop. The scheduler prioritizes programs based on their observational history such that any observations interrupted by weather or ToOs are rescheduled for subsequent nights. The WINTER scheduler divides the accessible northern sky ($\textgreater -36^{\circ}$ declination) into a fixed pointing grid of fields matching the WINTER field of view. Each potential observation is prioritized by a volumetric weighing metric and then scheduled with one of three modes: Gurobi-optimized observing, greedy observing, and queue observing. The volumetric metric maximizes the limiting volume of a requested field at a given altitude, which is based on detailed simulations of the limiting magnitude including seeing, sky brightness, and instrument noise. For nightly surveys, Gurobi-optimized observing utilizes the Gurobi linear optimizer to solve a traveling salesman problem, balancing various scientific surveys with unique cadences while minimizing overhead time from telescope slews. Alternatively, greedy observing continually selects the best field based on current volumetric weighting and is used to fill in reference sky images between scheduled science surveys and ToOs. Finally, for any ToO requests, queue observing sequentially steps through a predefined program. Figure \ref{fig:scheduler} shows WINTER observations from the first two years of observing, balancing various surveys as discussed in Section \ref{sec:performance}. 

\begin{figure*}[]
\epsscale{1}
\plotone{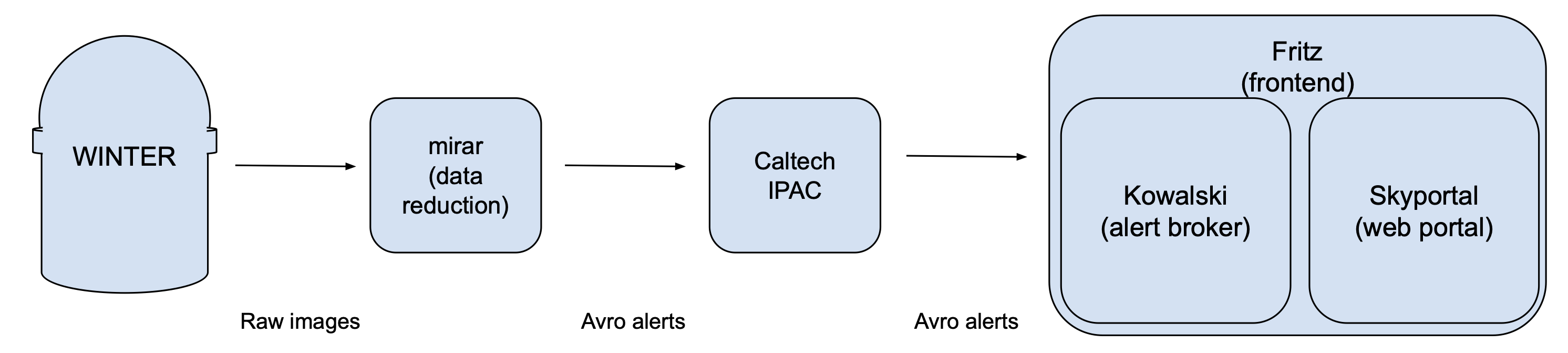}
\caption{A flow chart of the data reduction and alert pipelines, from the raw data collected by WINTER at Palomar Observatory, to the data reduction pipeline on a dedicated computer at Caltech, through daily transient candidate vetting on the Fritz web portal. 
\label{fig:drp_flow}}
\end{figure*}

For robotic follow up of gravitational-wave alerts, WINTER uses the Growth ToO Marshall \citep{GROWTH} to ingest alerts distributed via the LIGO-Virgo GCN system and produce optimized observing schedules with the gwemopt package \citep{Coughlin:2018lta, Coughlin:2019qkn, Ghosh:2015sxp}. \cite{Frostig:2022} outlines extensive simulations of WINTER observing with the gwemopt package, leading to a tailored gravitational-wave follow up strategy based on the reported distance and localization area of each gravitational-wave event. WINTER only follows up events with at least a 10$\%$ chance of kilonova detection based on simulations with two kilonova model grids. For each qualifying event, WINTER dynamically adjusts the image exposure time to optimize the search strategy, dedicating up to seven nights searching for each event and repeatedly observing kilonova candidates until they fade. 

\subsection{Data reduction pipeline} \label{sec:drp}
The WINTER data reduction and transient alert pipeline build upon the work for the ZTF and Palomar Gattini IR projects. The data reduction pipeline, named the Modular Image Reduction and Analysis Resource\footnote{https://github.com/winter-telescope/mirar} (\texttt{mirar}) is a open-source, modular, Python 3.11-based pipeline (V. R. Karambelkar and R. D. Stein et al., in prep). \texttt{mirar} processes both SUMMER and WINTER images, going through a series of standard infrared data reduction steps:

\begin{enumerate}
    \item Quality cuts: Rejection of bad images and masking of bad sensor regions.
    \item Calibration: Dark subtraction, flat-field correction, removal of sensor pattern noise through Fourier filtering, and subtraction of the modeled sky background with the SExtractor package \citep{Bertin1996}. 
    \item Astrometric calibration: Each image is astrometrically calibrated using astrometry.net \citep{astrometry.net} and a cross-matched catalogue of sources detected in Gaia \citep{Gaia} and 2MASS \citep{Cutri2003}. Bad images are once again rejected at this step.
    \item Stacking: Each WINTER image is split into a set of shorter exposures dithered around a central point, to enable long exposure times with the bright infrared sky background and reconstruction of the point spread function (PSF). The dithered exposures are stacked using the Swarp algorithm \citep{swarp}.
    \item Photometric calibration: Images are calibrated with Gaia, 2MASS, and Pan-STARRS. 
    \item Difference imaging: Nightly images are subtracted with the ZOGY package \citep{zogy} from UKIRT (J and Hs), Pan-STARRS (Y), or WINTER (Y, J, and Hs) reference images, where available. 
    \item Candidate identification: The SExtractor package is used to create a catalog of sources from the difference images, which allows from cross-matching, filtering, and resolved source naming. 
    \item Real-bogus classification: A machine-learning classifier distinguishes between image subtraction artifacts and real transient and variable sources.
\end{enumerate}

Sources classified as ``real'' are uploaded with the Apache Avro\footnote{https://avro.apache.org/} data serialization system to IPAC at Caltech, where they are sent to the Kowalski alert broker and the Skyportal frontend \citep{skyportal} for alert visualization and follow-up coordination (Figure \ref{fig:drp_flow}). Candidates appear daily on the Fritz web portal, which is a specific instance of Skyportal with a Kowalski backend, for human vetting and coordination of follow-up observations. 

\begin{figure}[]
\plotone{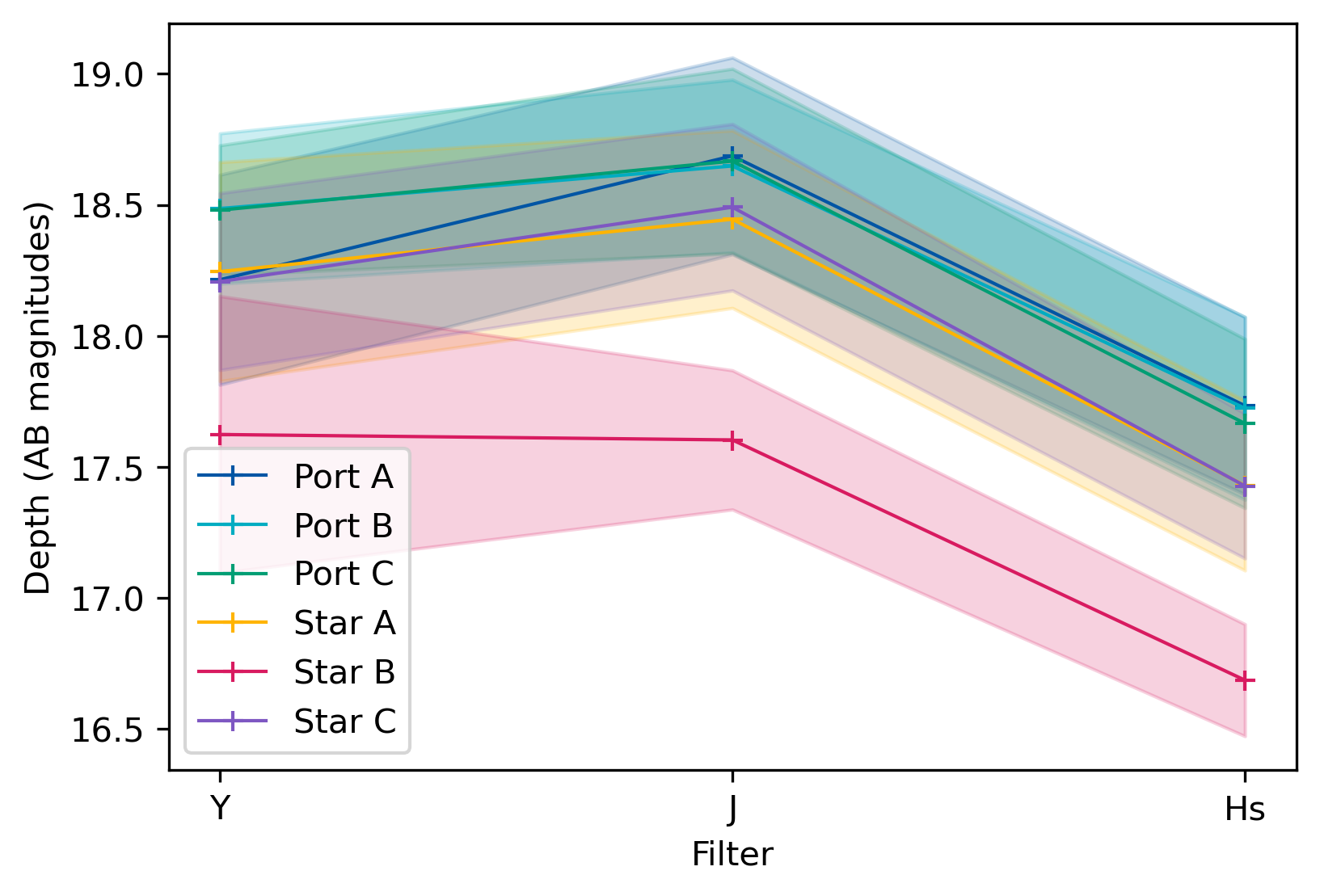}
\epsscale{1.2}
\caption{Depths of the six WINTER sensors during a period of Summer 2025 in the three project filters, $Y$, $J$, and $Hs$, with 16-minute exposure times for each band. Lines show the median depth with shaded regions indicating $\pm1\sigma$ uncertainty. The plot demonstrated the relative depths of filters, with a shallower response in $Hs$-band due to the increased sky background, and the spread of response from the six sensors. The data reduction pipeline has been optimized for $J$-band performance and not yet optimized for $Y$ or $Hs$. These differences in performance result from differences in dark current, gain, and quantum efficiency in each sensor.
\label{fig:depths}}
\end{figure}

\begin{figure*}[]
    \centering
    \gridline{\fig{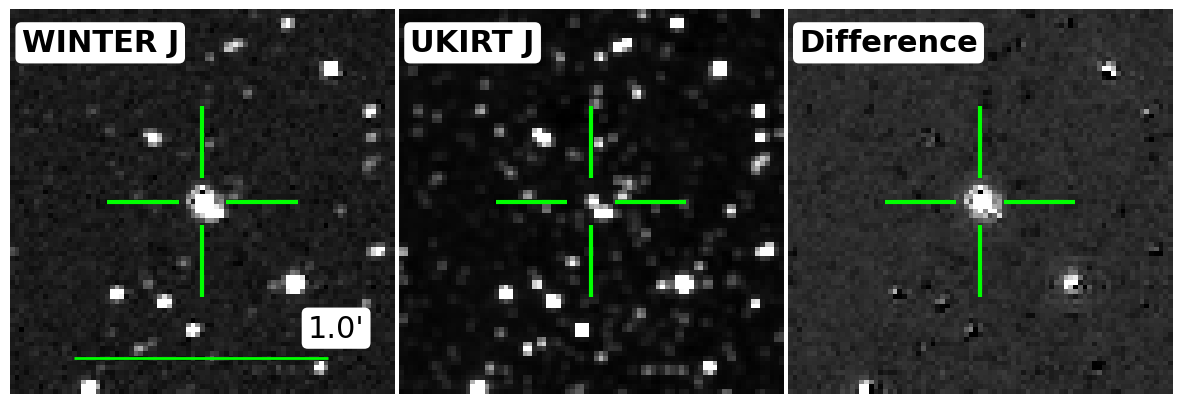}{0.48\textwidth}{}
              \fig{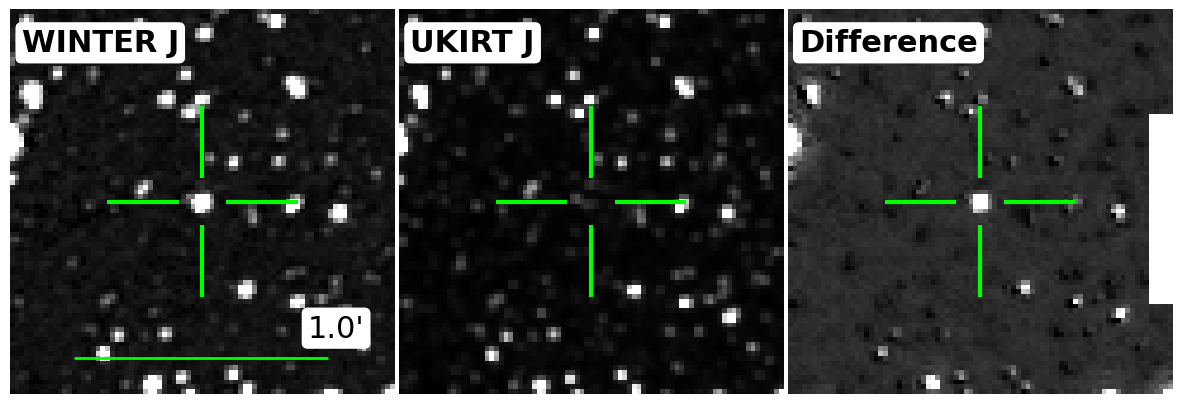}{0.48\textwidth}{}}
    \gridline{\fig{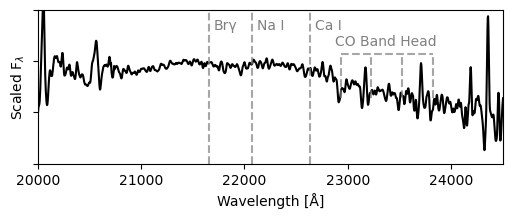}{0.46\textwidth}{}
              \fig{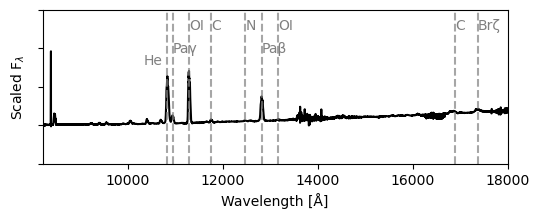}{0.49\textwidth}{}}
    \caption{Some examples of WINTER Galactic plane science. Left: WNTR24jheuy (Gaia18dvy/WTP17aadrne), a young stellar object outbursting (top) with follow-up MMIRs classification (bottom). Right: WNTR24fzjjq (ZTF24aaomlxy), a galactic classical nova with a WINTER detection (top) and FIRE classification (bottom).}
    \label{fig:plotfour}
\end{figure*}

\section{On-sky Performance and Early Science} \label{sec:performance}

WINTER has been on sky since June of 2023, running robotically to observe a mix of long-term surveys with ToO interrupts. The supervisory program, \texttt{wsp}, manages scheduling, sensor startup and shutdown, focusing, science observations, and dark frame collection. All operations are logged in real time on a dedicated \texttt{slack} channel, with alerts sent to relevant personnel. While a Caltech Optical Observatories member supervises each night, the system generally runs without human intervention. The data reduction pipeline processes the data in real time and candidate transient events are reviewed by members of the WINTER team.

\subsection{Photometric depth}
The original design for WINTER predicted a limiting magnitude of $J_{AB} = 21$ for a 5-minute exposure under ideal observing conditions \citep{Frostig:2020}. Multiple factors in the realization of the instrument led to the limiting magnitudes of WINTER to be $J_{AB} \sim 18.5$ in a 16-minute exposure. The most significant change is the effective $10\%$ QE, reduced from the designed $80\%$ QE (see Section \ref{sec:qe}). Additionally, increased dark current, higher gain, and a larger focused spot size compared to the original design introduce secondary performance degradations (Table \ref{tab:discrepancy}). There are also significant variations in performance between the six sensors, caused by differences in quantum efficiency, gain, dark current at the final operating temperatures, and hybridization issues present in some sensors (e.g., the glow spots seen in Figure \ref{fig:raw}). This leads to a median depth in a 16-minute exposure of 17.6-18.6 in $Y$ (8 dithers), 17.6-18.7 in $J$ (8 dithers), and 16.7-17.7 in $Hs$ (15 dithers) for the six sensors for MJD 60849 - 60856 (Figure \ref{fig:depths}). The data reduction pipeline is optimized around $J$-band imaging, with $Y$ and $Hs$ optimization planned. The depths also vary significantly with the conditions of the sky background.

\subsection{Early science with WINTER}
\begin{figure*}[]
\epsscale{0.75}
\plotone{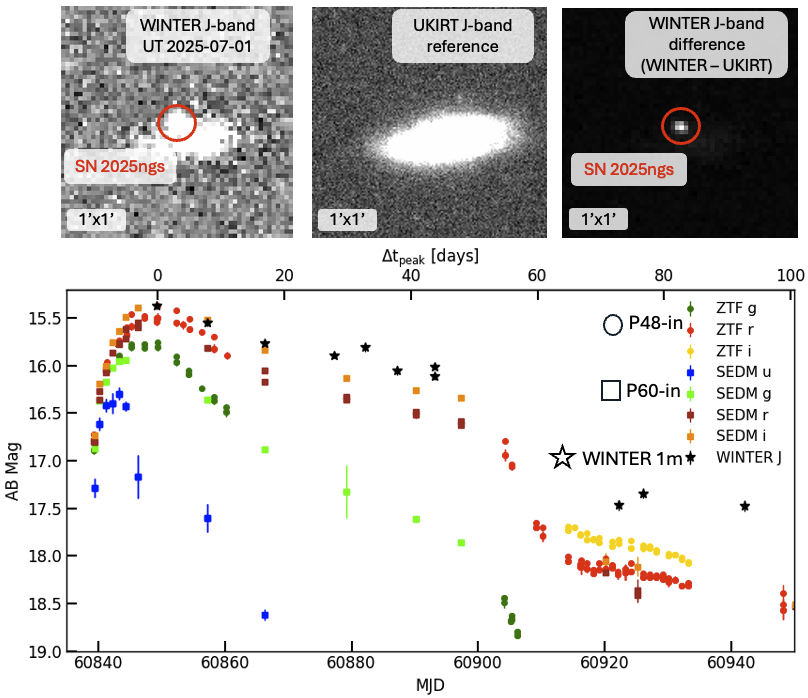}
\caption{An example of WINTER extragalactic science, showing WNTR25eetkd (SN 2025ngs), a type II supernova at 28.5 Mpc. Top: a cutout of the WINTER $J$-band image (left), a UKIRT reference image (middle), and the difference image (right) showing the transient. Bottom: A light curve of WNTR25eetkd in WINTER's $J$-band, compared with the light curve from optical surveys co-located at Palomar observatory.  
\label{fig:sn}}
\end{figure*}

WINTER observations balance a mix of long-term surveys with ToO interrupts (Figure \ref{fig:scheduler} shows the current history of WINTER observations). The core WINTER community surveys target regions of the sky predicted to yield the most infrared transients and variables in a blind search. This includes a 14-day cadence Galactic plane survey in $J$- and $Hs$-bands and a series of galaxy-targeted surveys. Galaxies within 25 Mpc, ULIRGS, and galaxy clusters are surveyed at a 10-day cadence in $J$-band. For the first year of operations, very nearby galaxies, such as M31, were monitored daily in $Y$-, $J$-, or $Hs$-bands. In addition to the core surveys, WINTER builds up $J$-band reference images of the sky, prioritizing fields above 60$^{\circ}$ declination where there are no UKIRT or VISTA images. The surveys are regularly interrupted by ToO requests with minutes-scale response times. The WINTER collaboration has published over seventy General Coordinates Network (GCN) circulars, including regular $J$-band detections of gamma-ray bursts (e.g., 
\citealt{2025GCN.42562....1M}) and detections of potential counterparts to neutrino alerts (e.g., \citealt{2024GCN.35584....1S}).

WINTER sets itself apart from other infrared surveys with its time-domain infrastructure, enabling rapid alerts and prompt spectroscopic follow-up of transient sources. For example, WINTER observed a young stellar object outburst (Figure \ref{fig:plotfour}). The team followed up the event with the MMIRS spectrograph on the MMT \citep{MMIRS, MMIRS_drp} and classified the event as an FU Ori type outburst following the classification scheme in \cite{Connelley:2018}. This event is also observed and studied in \citep{Gaia18dvy}. WINTER also observed a highly reddened galactic classical nova \citep{2024ATel16658....1W} (Figure \ref{fig:plotfour}). Follow-up spectroscopy with the FIRE spectrograph on the Magellan telescopes \citep{FIRE} shows the presence of carbon lines, which could indicate the event is a Fe II classical nova \citep{NIRnovae}. Other science with WINTER includes infrared coverage of supernovae, which are well covered in optical surveys but not infrared surveys (Figure \ref{fig:sn}), and luminous red novae in nearby galaxies. 

The collaboration has released three studies with early WINTER data. \citet{Frostig:2025} present near-IR follow-up of the GW candidate S250206dm, covering 43\% of the 50\% localization region but finding no viable kilonova candidates, consistent with the large inferred distance of $\sim$373\,Mpc. \citet{Karambelkar:2025} report the slow eruption of an early-AGB star in M31 (WNTR23bzdiq/WTP19aalzlk), which exhibited a decade-long optical/IR brightening and spectroscopic properties resembling stellar merger transients, possibly signaling the onset of common-envelope evolution. Finally, \citet{Frostig:2025b} characterizes two infrared FU~Ori outbursts discovered by WINTER and NEOWISE, including one new confirmation of an embedded Class~I protostar, demonstrating WINTER’s ability to uncover obscured protostellar eruptions inaccessible to optical surveys.

\section{Conclusions} \label{sec:conclusions}
WINTER has been running nightly robotic near-IR surveys since June 2023. Through a mix of extragalactic and Galactic plane surveys, targeted follow-up, and tiled searches, the project is enabling studies of reddened and intrinsically red transients and variables. The project's time-domain infrastructure supports rapid follow-up of new discoveries. Recent highlights from early WINTER science include searches for kilonovae from gravitational-wave triggers, evidence of common envelope evolution in AGB stars, and characterization of outbursts and mass accretion in young stellar objects. 

While WINTER's sensors did not meet design specifications, the project demonstrates the promise and need for wide-format InGaAs sensors at an accessible cost. Extensive testing presented here found decreased quantum efficiency from the design specification (from 80\% to $\sim$10\%) due to an issue in hybridization or processing of this lot of sensors. The InGaAs technology has no fundamental shortcomings, and the read out integrated circuit, and custom read out electronics all performed near expectations. The decreased quantum efficiency, along with secondary effects from increased dark current, gain, and focus spot size, reduced the on-sky depth by $\sim$2 magnitudes.  However, a broad range of science is underway at the present average depth of $J_{AB} = 18.5$ mag, which is significantly fainter than historical IR time-domain surveys.

WINTER studies complement and prepare for other infrared surveys. Future surveys such as the Roman Space Telescope and Cryoscope will enable wider and deeper near-IR observations, extending studies of these transients to greater depths. In the meantime, WINTER is developing near-IR infrastructure, surveys, and science for the time-domain era.

\section{Acknowledgments}
WINTER’s construction is made possible by the National Science Foundation under MRI grant number AST-1828470 with early operations supported by AST-1828470. Significant support for WINTER also comes from the California Institute of Technology, the Caltech Optical Observatories, the Bruno Rossi Fund of the MIT Kavli Institute for Astrophysics and Space Research, the David and Lucille Packard Foundation, and the MIT Department of Physics and School of Science. 

WINTER continued operation relies on the dedicated efforts of the Caltech Optical Observatories engineering and technical staff, support astronomers, and telescope operators at Palomar Observatory. Particularly, but not exclusively, Robert Sandoval, Gregord van Idsinga, Jon Derek Knapp, Drew Roderick, Isaac Wilson, Kathleen Koviac, and Paul Nied.

The team acknowledges the ongoing dedicated efforts of Stefan Lauxtermann and Per-Olov Pettersson and the detector teams at Sensor Creations/Attollo Engineering, throughout the detector production and commissioning, and Bryan Gall at Teledyne-FLIR.

Others at MIT and Caltech contributed to various aspects of the WINTER software and pipeline development, and analysis, including Benjamin Schneider, Tomas Ahumada, and Christopher Layden. Undergraduate and high school students contributed to the program with support from the Caltech SURF and MIT UROP programs: Allan Garcia-Zych, Joshua Glass, Sulekha Kishore, and Aswin Suresh.

Many others provided invaluable input to and review of various hardware and software subsystems.

D.F.'s contribution to this material is based upon work supported by the National Science Foundation under Award No. AST-2401779. This research award is partially funded by a generous gift of Charles Simonyi to the NSF Division of Astronomical Sciences. The award is made in recognition of significant contributions to Rubin Observatory’s Legacy Survey of Space and Time. 

This work made use of Astropy:\footnote{http://www.astropy.org} a community-developed core Python package and an ecosystem of tools and resources for astronomy \citep{astropy:2013, astropy:2018, astropy:2022}.

\bibliography{winter}{}
\bibliographystyle{aasjournal}

\end{document}